\documentclass[twocolumn,preprint]{aastex631}

\usepackage{amsmath,amstext,natbib,apjfonts,longtable,verbatim,float,graphicx,color, longtable, soul, enumitem} 

\newcommand{\msun}{M_\odot}

\newcommand{\HeI}{\hbox{{\rm He}~{\sc i}}}

\newcommand{\NII}{\hbox{{\rm [N}~{\sc ii}{\rm ]}}}

\newcommand{\OIII}{\hbox{{\rm [O}~{\sc iii}{\rm ]}}}

\newcommand{\Lya}{\hbox{{\rm Ly}$\alpha$}}
\newcommand{\Ha}{\hbox{{\rm H}$\alpha$}}
\newcommand{\Hb}{\hbox{{\rm H}$\beta$}}

\newcommand{\Pd}{\hbox{{\rm P}$\delta$}}

\shorttitle{The Rise of Faint, Red AGN at $z>4$}
\shortauthors{Kocevski et al.}
\graphicspath{{./}{figures/}}

\begin{document}

\title{\large \bf The Rise of Faint, Red AGN at $z>4$: \\ A Sample of Little Red Dots in the JWST Extragalactic Legacy Fields}

\suppressAffiliations

\author[0000-0002-8360-3880]{Dale D. Kocevski}
\affiliation{Department of Physics and Astronomy, Colby College, Waterville, ME 04901, USA}

\author[0000-0001-8519-1130]{Steven L. Finkelstein}
\affiliation{Department of Astronomy, The University of Texas at Austin, 2515 Speedway, Stop C1400, Austin, TX 78712, USA}

\author[0000-0001-6813-875X]{Guillermo Barro}
\affiliation{Department of Physics, University of the Pacific, Stockton, CA 90340 USA}

\author[0000-0003-1282-7454]{Anthony J. Taylor}
\affiliation{Department of Astronomy, The University of Texas at Austin, Austin, TX, USA}

\author[0000-0003-2536-1614]{Antonello Calabr{\`o}}
\affiliation{Osservatorio Astronomico di Roma, via Frascati 33, Monte Porzio Catone, Italy}

\author[0000-0001-9996-9732]{Brivael Laloux}
\affiliation{Centre for Extragalactic Astronomy, Department of Physics, Durham University, UK}
\affiliation{Institute for Astronomy \& Astrophysics, National Observatory of Athens, V. Paulou \& I. Metaxa, 11532, Greece}

\author[0000-0003-0426-6634]{Johannes Buchner}
\affiliation{Max Planck Institute for Extraterrestrial Physics, Giessenbachstrasse, 85741 Garching, Germany}

\author[0000-0002-1410-0470]{Jonathan R. Trump}
\affiliation{Department of Physics, 196 Auditorium Road, Unit 3046, University of Connecticut, Storrs, CT 06269, USA}

\author[0000-0002-9393-6507]{Gene C. K. Leung}
\affiliation{Department of Astronomy, The University of Texas at Austin, Austin, TX, USA}

\author[0000-0001-8835-7722]{Guang Yang}
\affiliation{Kapteyn Astronomical Institute, University of Groningen, P.O. Box 800, 9700 AV Groningen, The Netherlands}
\affiliation{SRON Netherlands Institute for Space Research, Postbus 800, 9700 AV Groningen, The Netherlands}

\author[0000-0001-5414-5131]{Mark Dickinson}
\affiliation{NSF's National Optical-Infrared Astronomy Research Laboratory, 950 N. Cherry Ave., Tucson, AZ 85719, USA}

\author[0000-0003-4528-5639]{Pablo G. P\'erez-Gonz\'alez}
\affiliation{Centro de Astrobiolog\'{\i}a (CAB), CSIC-INTA, Ctra. de Ajalvir km 4, Torrej\'on de Ardoz, E-28850, Madrid, Spain}

\author[0000-0001-9879-7780]{Fabio Pacucci}
\affiliation{Center for Astrophysics $\vert$ Harvard \& Smithsonian, Cambridge, MA 02138, USA}
\affiliation{Black Hole Initiative, Harvard University, Cambridge, MA 02138, USA}

\author[0000-0001-9840-4959]{Kohei Inayoshi}
\affiliation{Kavli Institute for Astronomy and Astrophysics, Peking University, Beijing 100871, China}

\author[0000-0002-6748-6821]{Rachel S. Somerville}
\affiliation{Center for Computational Astrophysics, Flatiron Institute, 162 5th Avenue, New York, NY 10010, USA}

\author[0000-0001-8688-2443]{Elizabeth J.\ McGrath}
\affiliation{Department of Physics and Astronomy, Colby College, Waterville, ME 04901, USA}


\author[0000-0003-3596-8794]{Hollis B. Akins}
\affiliation{Department of Astronomy, The University of Texas at Austin, 2515 Speedway Blvd Stop C1400, Austin, TX 78712, USA}


\author[0000-0002-9921-9218]{Micaela B. Bagley}
\affiliation{Department of Astronomy, The University of Texas at Austin, Austin, TX, USA}

\author[0000-0003-3917-1678]{Rebecca A.A. Bowler}
\affiliation{Jodrell Bank Centre for Astrophysics, Department of Physics and Astronomy, School of Natural Sciences, The University of Manchester, Manchester, M13 9PL, UK}

\author[0000-0003-0492-4924]{Laura Bisigello}
\affiliation{Dipartimento di Fisica e Astronomia "G.Galilei", Universit\'a di Padova, Via Marzolo 8, I-35131 Padova, Italy}
\affiliation{INAF--Osservatorio Astronomico di Padova, Vicolo dell'Osservatorio 5, I-35122, Padova, Italy}

\author[0000-0002-1482-5818]{Adam Carnall}
\affiliation{Institute for Astronomy, University of Edinburgh, Royal Observatory, Edinburgh, EH9 3HJ, UK}

\author[0000-0002-0930-6466]{Caitlin M. Casey}
\affiliation{Department of Astronomy, The University of Texas at Austin, Austin, TX, USA}

\author[0000-0001-8551-071X]{Yingjie Cheng}
\affiliation{University of Massachusetts Amherst, 710 North Pleasant Street, Amherst, MA 01003-9305, USA}

\author[0000-0001-7151-009X]{Nikko J. Cleri}
\affiliation{Department of Physics and Astronomy, Texas A\&M University, College Station, TX, 77843-4242 USA}
\affiliation{George P.\ and Cynthia Woods Mitchell Institute for Fundamental Physics and Astronomy, Texas A\&M University, College Station, TX, 77843-4242 USA}

\author[0000-0001-6820-0015]{Luca Costantin}
\affiliation{Centro de Astrobiolog\'{\i}a (CAB), CSIC-INTA, Ctra. de Ajalvir km 4, Torrej\'on de Ardoz, E-28850, Madrid, Spain}

\author[0000-0002-3736-476X]{Fergus Cullen}
\affiliation{Institute for Astronomy, University of Edinburgh, Royal Observatory, Edinburgh, EH9 3HJ, UK}
   
\author[0000-0001-8047-8351]{Kelcey Davis}
\altaffiliation{NSF Graduate Research Fellow}
\affiliation{Department of Physics, 196A Auditorium Road, Unit 3046, University of Connecticut, Storrs, CT 06269, USA}

\author[0000-0002-7622-0208]{Callum T. Donnan}
\affiliation{Institute for Astronomy, University of Edinburgh, Royal Observatory, Edinburgh, EH9 3HJ, UK}

\author{James S. Dunlop}
\affiliation{Institute for Astronomy, University of Edinburgh, Royal Observatory, Edinburgh, EH9 3HJ, UK}

\author[0000-0001-7782-7071]{{Richard S.} {Ellis}}
\affiliation{Jodrell Bank Centre for Astrophysics, Department of Physics and Astronomy, School of Natural Sciences, The University of Manchester, Manchester, M13 9PL, UK}

\author[0000-0001-7113-2738]{{Henry C.} {Ferguson}}
\affiliation{Space Telescope Science Institute, 3700 San Martin Drive, Baltimore, MD 21218, USA}

\author[0000-0001-7201-5066]{Seiji Fujimoto}\altaffiliation{Hubble Fellow}
\affiliation{Department of Astronomy, The University of Texas at Austin, Austin, TX 78712, USA}

\author[0000-0003-3820-2823]{{Adriano} {Fontana}}
\affiliation{INAF - Osservatorio Astronomico di Roma, via di Frascati 33, 00078 Monte Porzio Catone, Italy}

\author[0000-0002-7831-8751]{{Mauro} {Giavalisco}}
\affiliation{University of Massachusetts Amherst, 710 North Pleasant Street, Amherst, MA 01003-9305, USA}

\author[0000-0002-5688-0663]{Andrea Grazian}
\affil{INAF--Osservatorio Astronomico di Padova, 
Vicolo dell'Osservatorio 5, I-35122, Padova, Italy\\}

\author[0000-0001-9440-8872]{Norman A. Grogin}
\affiliation{Space Telescope Science Institute, Baltimore, MD, USA}

\author[0000-0001-6145-5090]{Nimish P. Hathi}
\affiliation{Space Telescope Science Institute, 3700 San Martin Drive, Baltimore, MD 21218, USA}

\author[0000-0002-3301-3321]{Michaela Hirschmann}
\affiliation{Institute of Physics, Laboratory of Galaxy Evolution, Ecole Polytechnique Fédérale de Lausanne (EPFL), Observatoire de Sauverny, 1290 Versoix, Switzerland}

\author[0000-0002-1416-8483]{Marc Huertas-Company}
\affiliation{Instituto de Astrofísica de Canarias (IAC), La Laguna, E-38205,
Spain}
\affiliation{Observatoire de Paris, LERMA, PSL University, 61 avenue de
l’Observatoire, F-75014 Paris, France}
\affiliation{Université Paris-Cité, 5 Rue Thomas Mann, 75014 Paris, France}
\affiliation{Universidad de La Laguna. Avda. Astrofísico Fco. Sanchez, La Laguna, Tenerife, Spain}
\affiliation{Center for Computational Astrophysics, Flatiron Institute, New York, NY 10010, USA}

\author[0000-0002-4884-6756]{Benne W. Holwerda}
\affiliation{University of Louisville, Department of Physics and Astronomy, 102 Natural Science Building, Louisville, KY 40292, USA}

\author[0000-0002-8096-2837]{Garth Illingworth}
\affiliation{Department of Astronomy and Astrophysics, UCO/Lick Observatory, University of California, Santa Cruz, CA 95064, USA}

\author[0000-0002-0000-2394]{St{\'e}phanie Juneau}
\affiliation{NSF's NOIRLab, 950 N. Cherry Ave., Tucson, AZ 85719, USA}

\author[0000-0001-9187-3605]{Jeyhan S. Kartaltepe}
\affiliation{Laboratory for Multiwavelength Astrophysics, School of Physics and Astronomy, Rochester Institute of Technology, 84 Lomb Memorial Drive, Rochester, NY 14623, USA}

\author[0000-0002-6610-2048]{Anton M. Koekemoer}
\affiliation{Space Telescope Science Institute, 3700 San Martin Dr., Baltimore, MD 21218, USA}

\author[0000-0002-1044-4081]{Wenxiu Li}
\affiliation{Kavli Institute for Astronomy and Astrophysics, Peking University, Beijing 100871, China}

\author[0000-0003-1581-7825]{Ray A. Lucas}
\affiliation{Space Telescope Science Institute, 3700 San Martin Drive, Baltimore, MD 21218, USA}

\author[0000-0002-6668-2011]{Dan Magee}
\affiliation{Department of Astronomy and Astrophysics, UCO/Lick Observatory, University of California, Santa Cruz, CA 95064, USA}

\author{Charlotte Mason}
\affiliation{Cosmic Dawn Center (DAWN), Jagtvej 128, DK-2200, Copenhagen N, Denmark}
\affiliation{Niels Bohr Institute, University of Copenhagen, Jagtvej 128, DK-2200, Copenhagen N, Denmark}

\author[0000-0003-4368-3326]{Derek J. McLeod}
\affiliation{Institute for Astronomy, University of Edinburgh, Royal Observatory, Edinburgh, EH9 3HJ, UK}

\author{Ross J. McLure}
\affiliation{Institute for Astronomy, University of Edinburgh, Royal Observatory, Edinburgh, EH9 3HJ, UK}

\author[0000-0002-8951-4408]{Lorenzo Napolitano}
\affiliation{INAF – Osservatorio Astronomico di Roma, via Frascati 33, 00078, Monteporzio Catone, Italy}
\affiliation{Dipartimento di Fisica, Università di Roma Sapienza, Città Universitaria di Roma - Sapienza, Piazzale Aldo Moro, 2, 00185, Roma, Italy}

\author[0000-0001-7503-8482]{Casey Papovich}
\affiliation{Department of Physics and Astronomy, Texas A\&M University, College Station, TX, 77843-4242 USA}
\affiliation{George P.\ and Cynthia Woods Mitchell Institute for Fundamental Physics and Astronomy, Texas A\&M University, College Station, TX, 77843-4242 USA}

\author[0000-0003-3382-5941]{Nor Pirzkal}
\affiliation{ESA/AURA Space Telescope Science Institute}

\author[0000-0002-9415-2296]{Giulia Rodighiero}
\affiliation{Department of Physics and Astronomy, Università degli Studi di Padova, Vicolo dell’Osservatorio 3, I-35122, Padova, Italy}
\affiliation{INAF - Osservatorio Astronomico di Padova, Vicolo dell’Osservatorio 5, I-35122, Padova, Italy}

\author[0000-0002-9334-8705]{Paola Santini}
\affiliation{INAF - Osservatorio Astronomico di Roma, via di Frascati 33, 00078 Monte Porzio Catone, Italy}

\author[0000-0003-3903-6935]{Stephen M.~Wilkins} %
\affiliation{Astronomy Centre, University of Sussex, Falmer, Brighton BN1 9QH, UK}
\affiliation{Institute of Space Sciences and Astronomy, University of Malta, Msida MSD 2080, Malta}

\author[0000-0003-3466-035X]{{L. Y. Aaron} {Yung}}
\affiliation{Space Telescope Science Institute, 3700 San Martin Drive, Baltimore, MD 21218, USA}

\begin{abstract}

We present a sample of 341 "little red dots" (LRDs) spanning the redshift range $z\sim2-11$ using data from the CEERS, PRIMER, JADES, UNCOVER and NGDEEP surveys. Unlike past use of color indices to identify LRDs, we employ continuum slope fitting using shifting bandpasses to sample the same rest-frame emission blueward and redward of the Balmer break. This enables the detection of LRDs over a wider redshift range and with less contamination from galaxies with strong breaks that otherwise lack a rising red continuum. The redshift distribution of our sample increases at $z<8$ and then undergoes a rapid decline at $z\sim4.5$, which may tie the emergence of these sources to the inside-out growth that galaxies experience during this epoch. We find that LRDs are $\sim1$ dex more numerous than X-ray and UV selected AGN at z~5-7.  Within our sample, we have identified the first two X-ray detected LRDs. An X-ray spectral analysis confirms that these AGN are moderately obscured with $\log\,(N_{\rm H}/{\rm cm}^{2}$) of $23.3^{+0.4}_{-1.3}$ and $22.72^{+0.13}_{-0.16}$. Our analysis reveals that reddened AGN emission dominates their rest-optical light, while the rest-UV originates from their host galaxies. We also present NIRSpec observations from the RUBIES survey of 17 LRDs that show broad emission lines consistent with AGN activity. The confirmed AGN fraction of our sample is 71\% for sources with F444W<26.5. In addition, we find three LRDs with blue-shifted Balmer absorption features in their spectra, suggesting an outflow of high-density, low-ionization gas from near the central engine of these faint, red AGN.
\vspace{-1in}

\end{abstract}


\section{Introduction}

One of the more surprising results to come from the first year of \textit{JWST} observations is the detection of numerous faint, broad-line AGN at $z>5$ \citep{onoue23,Kocevski23b, Harikane23, Matthee_2023, Maiolino23, Larson23, Greene_2023}.  These sources have luminosities that are 2-3 dex below that of bright quasars identified by ground-based surveys at similar redshifts (e.g., \citealt{Willott10b, jiang16,Mazzucchelli_2017,matsuoka22,Yang23}) and are powered by BHs with masses of $\approx 10^{6-7}~{\rm M_{\odot}}$, making them among the least-massive BHs known in the early Universe.  These faint quasars are more representative of the underlying BH population at high redshifts and potentially the key to constraining models of BH seeding \citep{Pacucci_2022_search, Li_Inayoshi23}, the contribution of AGN to hydrogen reionization \citep{Dayal20, Yung21, giallongo19,Finkelstein19}, and the early coevolution of galaxies and BHs \citep{Habouzit22, Inayoshi_2022,Pacucci_2023_overmassive}.

About 20\% of the broad-line AGN identified with \textit{JWST} appear to be heavily obscured, featuring a steep red continuum in the rest-frame optical, while also exhibiting relatively blue colors in the rest-frame UV \citep{Kocevski23b, Harikane23, Furtak24, Matthee_2023,Greene_2023,Killi23}.   Sources with this “v-shaped”, red plus blue spectral energy distribution (SED) have come to be known as “little red dots” (LRDs) in the literature \citep{Matthee_2023}\footnote{While the definition of a "little red dot" has varied in the literature, \newline in this study, the term is used to refer to compact sources with \newline red-optical and blue-UV colors.}.  Due to their unique colors and compact morphologies, LRDs have been identified photometrically and found to be quite ubiquitous, having number densities of $\sim10^{-5}$ Mpc$^{-3}$ mag$^{-1}$, which amounts to a few percent of the galaxy population at redshift $z\sim5-6$ \citep{Barro23, Labbe23, Kokorev24}.  Using follow-up spectroscopy, \citet{Greene_2023} demonstrated that over 80\% of photometrically-selected LRDs show broad-line emission when care is taken to exclude brown dwarf contaminants (e.g., \citealt{Langeroodi_2023}).

The origin of the red and blue emission in these sources has been heavily debated.  Their steep rest-frame optical slopes are consistent with either a reddened AGN continuum or emission from dusty star-formation \citep{Kocevski23b, Barro23, Labbe23, Akins23}, with lines of evidence supporting both scenarios.  For example, if the continuum is dominated by stellar emission, then the implied H$\alpha$ equivalent widths (EWs) would far exceed that of typical star-forming galaxies at lower redshift (i.e., \citealt{Greene_2023, Fumagalli12, Whitaker14}).  On the other hand, photometric constraints in the mid-infrared favor SED models consistent with a dusty, compact starburst and only mild contribution from an obscured AGN \citep{Williams23, Pablo24}. 

Several emission mechanisms have also been proposed to explain the UV excess observed in LRDs, including scattered light from the central AGN and stellar emission emerging from a relatively dust-free host or escaping unattenuated due to patchy dust \citep{Kocevski23b, Barro23, Labbe23, Akins23, Killi23}. While red sources with a UV excess have been found at lower redshifts in sources such as the hot dust-obscured galaxy (Hot DOGs) population \citep{Assef20}, they are rarer, making up only 1\% of IR-bright, dust-obscured galaxies \citep{Noboriguchi19}.  

A substantial portion of black hole (BH) growth in the Universe is thought to be heavily obscured \citep{Gilli07, Ueda14} and there are signs that the fraction of obscured AGN increases with redshift and decreasing luminosity \citep[][although see \citealt{Scholtz23}]{Hasinger08, Merloni14, Aird15, Vito18, Ni20, Peca_2023}.  If most LRDs are indeed powered by buried accretion, they would be an important population of previously-hidden AGN at high redshifts and provide a unique window into the early, obscured growth phase of today's supermassive black holes.

\begin{table}
\renewcommand\thetable{1} 
\caption{NIRCam Survey Areas and $5\sigma$ Point-Source Sensitivities}\vspace{0mm}
\begin{center}
\vspace{-0.1in}
\begin{tabular}{cccc}
\hline
\hline
Survey & Field & Area        & F444W   \\
       &       & (armin$^2$) &  (AB mag)\\
\hline
NGDEEP &  GDS & 11.4  & 30.65  \\ 
UNCOVER &  Abell 2744 & 43.8 & 29.33 \\
JADES  &  GDS & 62.1 & 29.62 \\ 
CEERS  &  EGS & 88.6 & 28.97 \\
PRIMER &  COS & 138.9 & 28.78 \\
PRIMER &  UDS & 243.0 &  28.54 \\ 
\hline 
\end{tabular}
\label{tab_obs}
\end{center}
 \tablecomments{Field acronyms: EGS = Extended Groth Strip; GDS = Great Observatories Origins Deep Survey - South (GOODS-S); UDS = UKIDSS Ultra Deep Survey; COS: Cosmic Evolution Survey (COSMOS)}
\end{table}

In this study, we construct the largest sample of photometrically-selected LRDs to date using publicly available \textit{JWST} imaging in several extragalactic legacy fields.  Our selection technique centers around measuring rest-frame UV and optical continuum slopes using shifting bandpasses in order to sample the same rest-frame emission blueward and redward of the Balmer break.  This enables us to self-consistently search for compact sources with red optical and blue UV colors over a wide range of redshifts ($z\sim2-11$).  It also allows us to identify lower-redshift analogs to the LRDs found at $z\sim6$, where the host galaxy may be more readily visible, helping us to understand the origin of their unusual SEDs. 

Our analysis is presented as follows. In Section 2 we describe the {\it JWST} imaging used for this study, while Section 3 describes our target selection and the details of our SED modeling.  We present our results in Section 4 and the implications of our findings are discussed in Section 5.  When necessary, the following cosmological parameters are used: $H_{0} = 70~{\rm km~s^{-1}~Mpc^{-1}; \Omega_{tot}, \Omega_{\Lambda}, \Omega_{m} = 1, 0.7, 0.3}$. 

\section{Data Description \& Reduction} \label{sec:obs_data} 

\subsection{NIRCam Imaging}

In this study, we make use of {\it JWST} NIRCam imaging from the Cosmic Evolution Early Release Science Survey (CEERS; \citet{finkelstein22}), the Next Generation Deep Extragalactic Exploratory Public (NGDEEP) survey \citep{Bagley23}, the JWST Advanced Deep Extragalactic Survey (JADES; Eisenstein et al.~2023), the Public Release IMaging for Extragalactic Research (PRIMER) survey \citep{Dunlop21}, and the Ultradeep NIRSpec and NIRCam Observations Before the Epoch of Reionization (UNCOVER) survey \citep{Bezanson22}.  The fields and area covered with NIRCam by each survey are listed in Table \ref{tab_obs}.

For the JADES survey, we make use of the public NIRCam mosaics made available as part of the JADES second data release\footnote{https://archive.stsci.edu/hlsp/jades}.  For UNCOVER, we use the reductions made available by the GLASS-JWST data release\footnote{https://archive.stsci.edu/hlsp/glass-jwst}.
The CEERS, PRIMER, and NGDEEP NIRCam data were processed using the {\it JWST} Calibration Pipeline\footnote{\url{http://jwst-pipeline.readthedocs.io/en/latest/}} (versions 1.8.5, 1.10.2, and 1.9.2, respectively) with custom modifications described in  \citet{finkelstein22} and \citet{bagley22}. The resulting images were registered to the same World Coordinate System reference frame (based on Gaia DR1.2; \citet{Gaia16}) and combined into a single mosaic for each field using the drizzle algorithm with an inverse variance map weighting \citep{fruchter02,Casertano00} via the Resample step in the pipeline.  The final mosaics in all fields have pixel scales of 0\farcs03/pixel.

Source detection and photometry on the NIRCam mosaics were computed on PSF-matched images using SExtractor \citep{bertin96} version~2.25.0 in two-image mode, with an inverse-variance weighted combination of the PSF-matched F277W and F356W images as the detection image.  Photometry was measured in all of the available NIRCam bands in each field, as well as the F606W and F814W HST bands using public data from the CANDELS and 3D-HST surveys \citep{grogin11, koekemoer11,brammer12,Momcheva16}.  This process is similar to that described in \citet{Finkelstein23}, so we refer the reader there for additional details.  

 \begin{table}
\caption{Bands Used to Derive Continuum Slope versus Redshift}\vspace{0mm}
\begin{center}
\vspace{-0.2in}
\begin{tabular}{ccc}
\hline
\hline
Redshift & $\beta_{\rm UV}$ & $\beta_{\rm opt}$   \\
\hline
{\footnotesize $2<z<3.25$}    &  {\footnotesize F606W, F814W, F115W} & {\footnotesize F150W, F200W, F277W} \\  
{\footnotesize $3.25<z<4.75$} &  {\footnotesize F814W, F115W, F150W} & {\footnotesize F200W, F277W, F356W} \\  
{\footnotesize $4.75<z<8$}    &  {\footnotesize F115W, F150W, F200W} & {\footnotesize F277W, F356W, F444W} \\  
{\footnotesize $z>8$}         &  {\footnotesize F150W, F200W, F277W} & {\footnotesize F356W, F444W} \\  
\hline 
\end{tabular}
\label{tab_fitting}
\end{center}
\vspace{-0.2in}
\end{table}

We compute photometric redshifts using the \textsc{eazy} \citep{brammer08} software package for all sources in our photometric catalogs.  \textsc{eazy} fits non-negative linear combinations of user-supplied templates to derive probability distribution functions (PDFs) for the redshift, based on the quality of fit of the various template combinations to the observed photometry for a given source.  We use the default template set ``tweak\_fsps\_QSF\_12\_v3", which consists of 12 templates derived from the stellar population synthesis code FSPS \citep{Conroy10}, as well as the six bluer templates created by \cite{Larson22}, as described in \cite{Finkelstein23}.  A flat redshift prior with respect to luminosity was assumed and redshifts from $z = 0$ to 20 were considered.  We also perform a “low-redshift” run with the maximum redshift set to $z = 7$ to allow for visualization of the best-fitting low-redshift model for sources found to be at $z>9$.

\subsection{NIRSpec Spectroscopy}

The NIRSpec observations used in this study consist of three pointings in the UDS field taken on 16, 18, and 19 January 2024 and six pointings in the EGS field taken on 13 and 20 March 2024 as part of the RUBIES program (GO-4233; PI: A. de Graaff).  The observations were taken with the G395M/F290LP $R\simeq 1000$ grating/filter pair using a three-nod pattern.  The coadded spectra have a total exposure time of 5689.7~s.  The data were reduced using version~1.13.4 of the JWST Science Calibration Pipeline with the Calibration Reference Data System (CRDS) mapping 1215, starting from the Level~0 uncalibrated data products (``\_uncal.fits'' files) available on MAST. Custom parameters were used for the \texttt{jump} step at the detector-level calibration for a better treatment of the ``snowballs''\footnote{\url{https://jwst-docs.stsci.edu/data-artifacts-and-features/snowballs-and-shower-artifacts}} produced by high-energy cosmic ray events, and a nodded background subtraction was adopted.

\begin{figure}
\centering
\includegraphics[width=\linewidth]{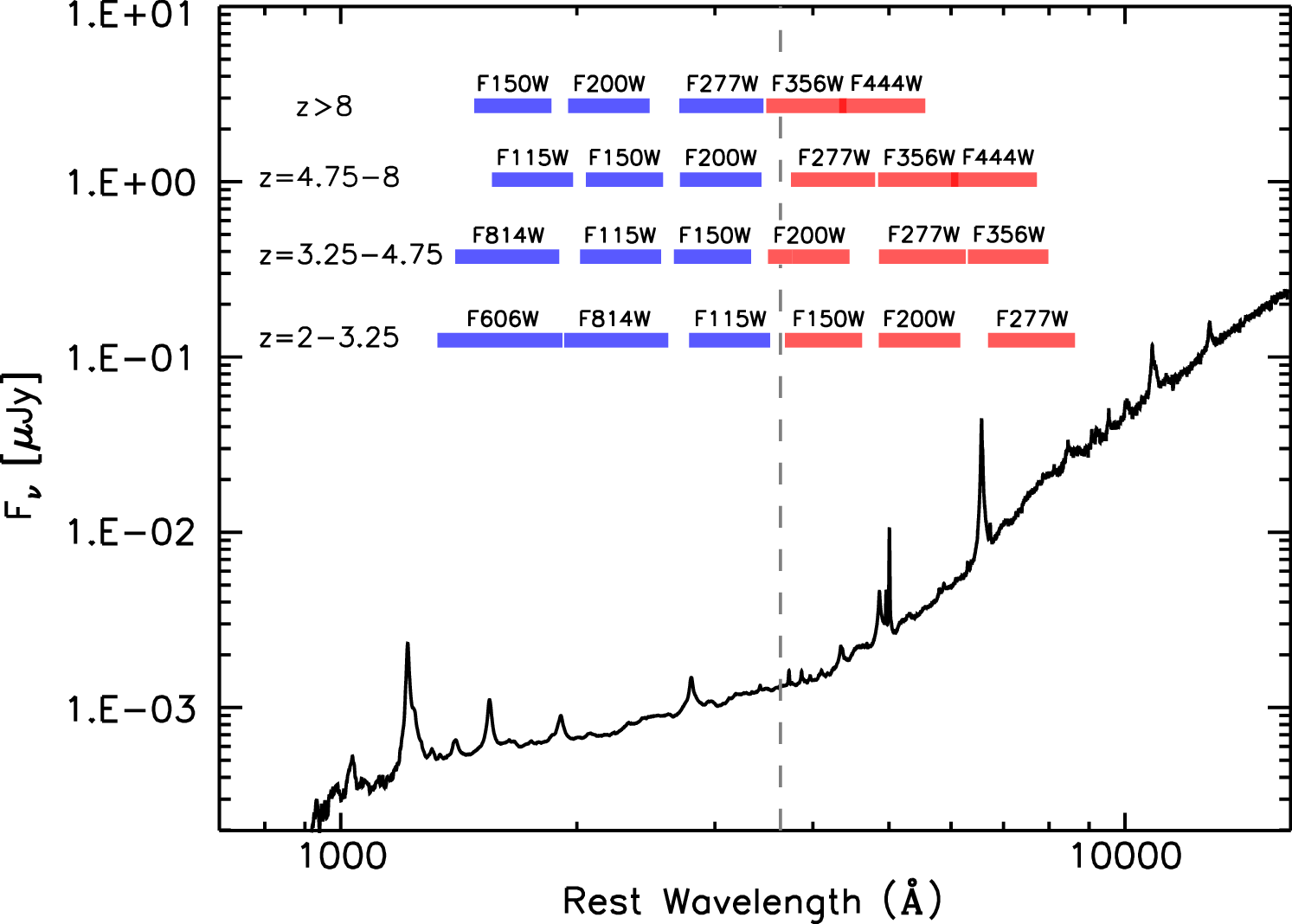}
\caption{Best-fit SED of the obscured, broad-line AGN CEERS 746 from \citet{Kocevski23b}.  The blue and red bars denote the filters blueward and redward of the Balmer break at 3645\AA~(the dashed vertical line) used to determine the rest-frame UV and optical continuum slopes, respectively, of each source given its redshift.}
\label{fig:QSO_SED_wFilters}
\end{figure}

\begin{figure*}
\centering
\includegraphics[width=\linewidth]{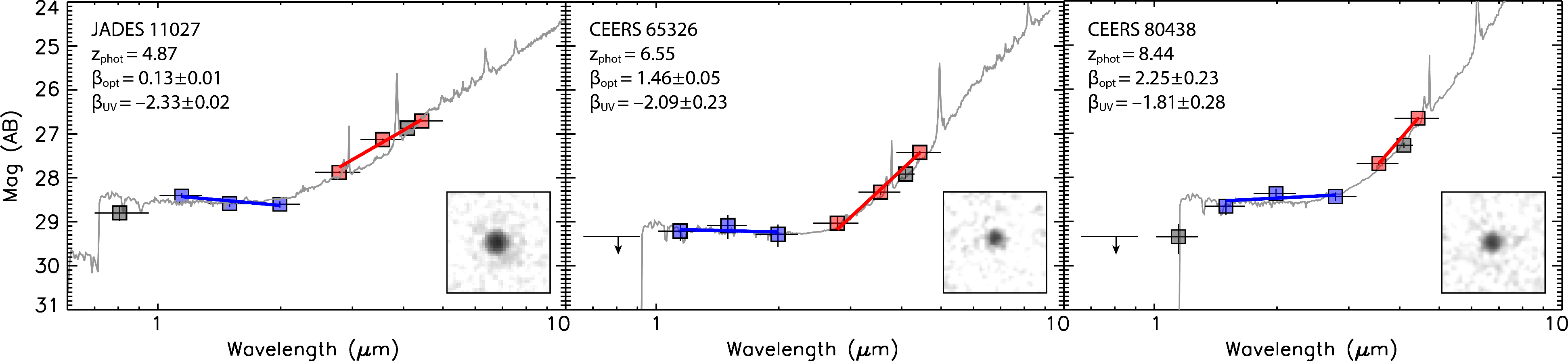}
\caption{Examples of our continuum slope fits for sources at a range of redshifts.  The bands used to measure the rest-frame UV and optical slopes are shown as blue and red squares, respectively.  The F444W image cutouts of each source are $1^{\prime\prime}\times1^{\prime\prime}$ in size.  The best-fit SEDs shown in light grey are galaxy plus QSO hybrid models (see \S3 for details). $2\sigma$ upper limits are shown for bands with non-detections.}
\label{fig:example_fits}
\end{figure*}

\begin{figure*}
\centering
\includegraphics[width=\linewidth]
{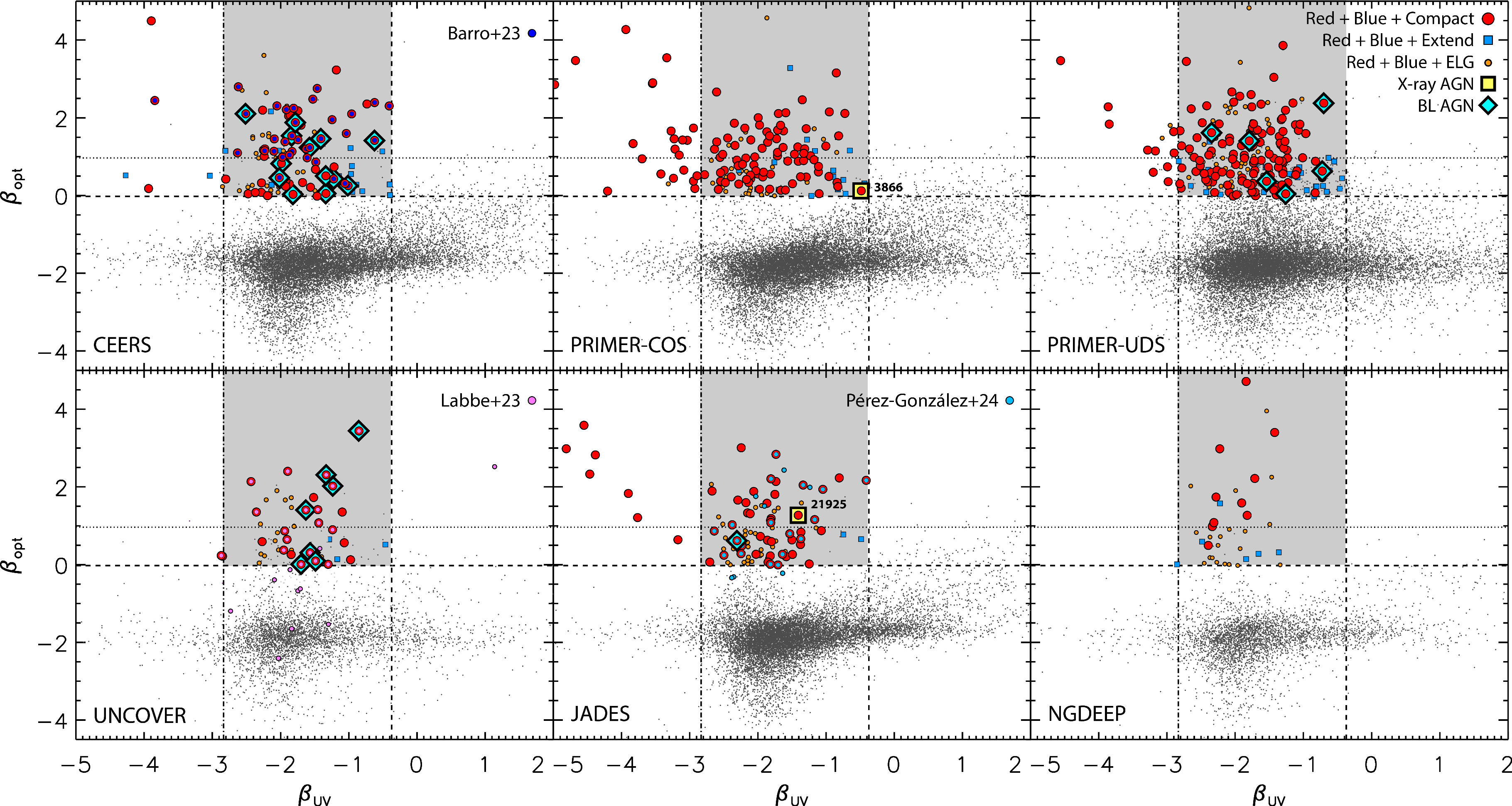}
\caption{Distribution of best-fit optical and UV spectral slopes, $\beta_{\rm opt}$ and $\beta_{\rm UV}$, measured in the CEERS, PRIMER-COSMOS, PRIMER-UDS, UNCOVER, JADES, and NGDEEP datasets for galaxies at $z>2$ detected in the F444W filter with a ${\rm SNR}>12$.  The horizontal and vertical dashed lines denote our selection criteria of $\beta_{\rm opt}>0$ and $\beta_{\rm UV}<-0.37$ meant to select sources with red and blue colors in the rest-frame optical and UV, respectively.  The horizontal dotted line denotes the $\beta_{\rm opt}$ limit that corresponds to the color cut used in \citet{Barro23} (i.e., ${\rm F277W-F444W} > 1.5$).  The vertical dot-dashed line denotes the $\beta_{\rm UV}$ limit that corresponds to the blue color cut used by \citet{Greene_2023} to exclude brown dwarfs (i.e., ${\rm F115W-F200W} < -0.5$).  The red circles are sources that satisfy both our spectral slope and size cuts (see \S3 for details). Also shown are sources excluded from our primary sample due to either failing our size cut (blue squares) or being flagged as potential strong line emitters (orange circles). The smaller light blue, dark blue, and pink circles are the sample of LRDs identified in \citet{Pablo24},  \citet{Barro23}, and \citet{Labbe23}, respectively.  The light blue diamonds indicate sources in our sample with broad emission line detections.
}
\label{fig:slopes}
\end{figure*}

The reduced two-dimensional (2D) spectra (``s2d'') have a rectified trace with a flat slope. To best optimize the extraction of one-dimensional (1D) spectra from the 2D spectra, we perform a weighted extraction based on the methodology of \cite{Horne86}. Briefly, for a given spectrum, we take the median of the 2D spectrum along the spectral direction to produce a spatial profile for the source. We then identity the central peak of this profile, which corresponds to the source's spectral trace. We then set all pixels in the spatial profile that are not a part of this central feature to zero and normalize the area under this masked spatial profile to one. We then use the normalized profile as the variable \textbf{P} in Table~1 of \cite{Horne86} and follow the prescription given therein to extract an optimized 1D spectrum.  

\section{Methodology}\label{sec:method}

\subsection{Sample Selection}\label{sec:sample}

Following the approach of \citet{Barro23} and \citet{Labbe23}, we identify LRDs as sources with a compact morphology that are red at rest-frame optical wavelengths and blue in the rest-frame UV.  While these previous studies used color indices measured in a single combination of bands (i.e., F277-F444W), in this study we perform our search using cuts on the UV and optical continuum slope measured by fitting photometry in multiple bands blueward and redward of the Balmer break at 3645\AA. 

To determine the continuum slope, $\beta$ (where $\beta$ is defined such that $f_{\lambda}\propto\lambda^{\beta})$, for a source, we perform a $\chi^{2}$ minimization fit to the observed magnitudes using the linear relationship: 
\begin{equation}
   m_{i} = -2.5~(\beta+2)~\log {(\lambda_{i})}+c 
\end{equation}
where $m_{i}$ is the AB magnitude measured in the $i$th filter with an effective wavelength of $\lambda_{i}$. We perform this fit to determine both the rest-frame UV and optical spectral slopes, $\beta_{\rm UV}$ and $\beta_{\rm opt}$, using fluxes measured in bands blueward and redward of 3645\AA~based on the redshift of each source.  The filter combinations used for these fits as a function of source redshift are listed in Table \ref{tab_fitting} and shown in Figure \ref{fig:QSO_SED_wFilters}.  Three bands are used to measure the slope in all cases except for sources at $z>8$, where only the F356W and F444W bands are used to determine the rest-frame optical slope.  Examples of our slope fits for sources at a range of redshifts are shown in Figure \ref{fig:example_fits}.

\begin{figure*}[h]
\centering
\includegraphics[width=\linewidth]{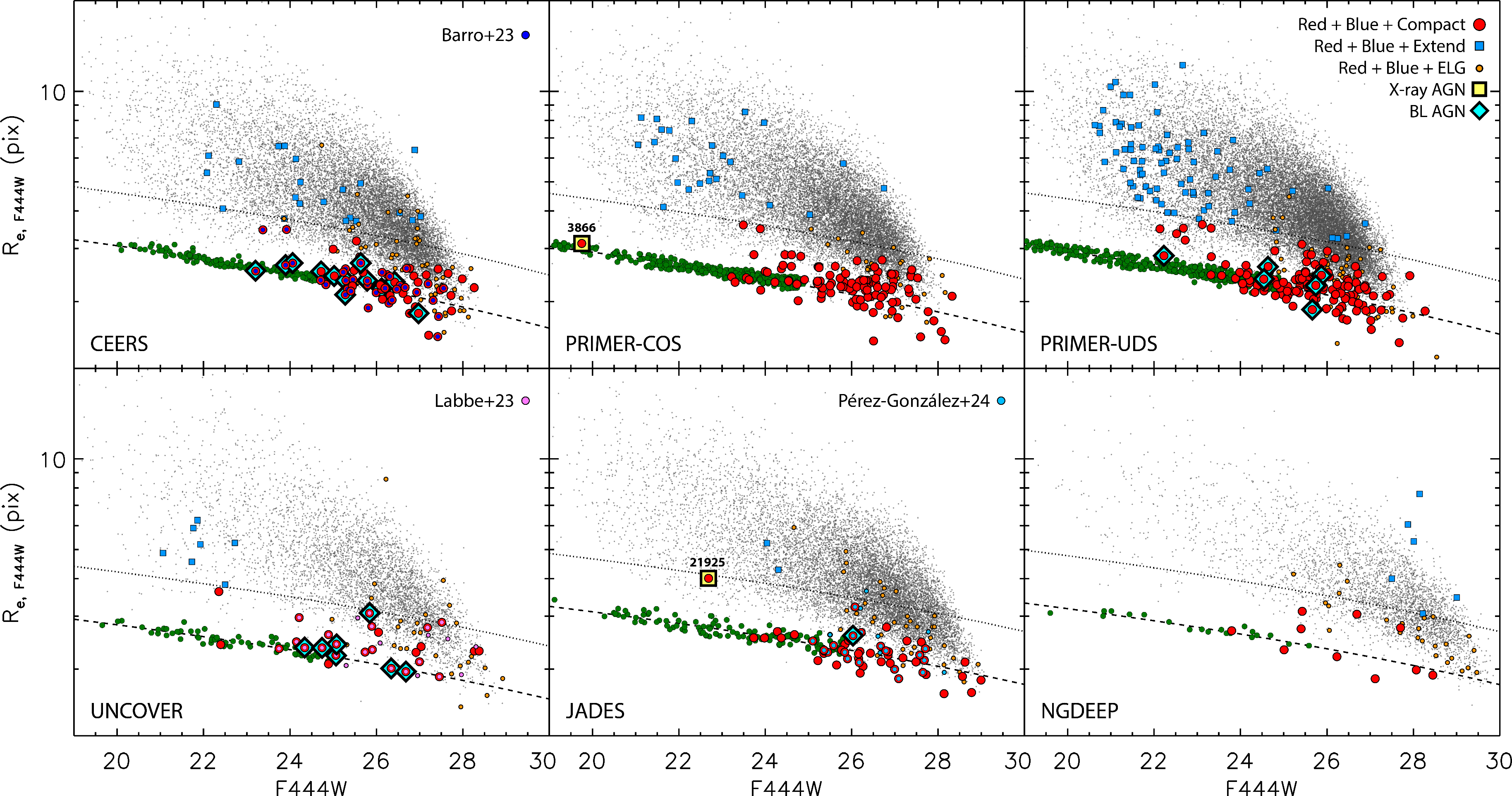}
\caption{F444W magnitude versus half-light radius for galaxies in the CEERS, PRIMER-COSMOS, PRIMER-UDS, UNCOVER, JADES, and NGDEEP datasets at $z>2$ detected in the F444W filter with a ${\rm SNR}>12$.  Green circles are stars and the dashed line denotes our best-fit to the stellar locus.  Our magnitude-dependant size cut is shown by the dotted line, which is 1.5 times our best-fit to the magnitude-size relationship for stars in each field.  Symbols are the same as in Figure \ref{fig:slopes}}
\label{fig:mag-size}
\end{figure*}

Uncertainties on our measured $\beta$ slopes are calculated using a Monte Carlo approach.  For each band, we perform 1000 random draws from a Gaussian whose mean is set to the measured flux in that band and whose standard deviation is set to the photometric error.  Our continuum slope measurements are then repeated on all 1000 mock SEDs and standard deviations are calculated from the resulting distributions.

We can translate the color cuts used by previous studies to identify extremely red objects into continuum slopes using the relationship
\begin{equation}
    \beta = \frac{0.4(m_{1}-m_{2})_{\rm AB}}{\log\,(\lambda_{2}/\lambda_{1})}-2
\label{eq:color-slope-conv} 
\end{equation}

For this study, we select red sources with a UV-excess using an optical continuum slope cut of $\beta_{\rm opt}>0$  and a UV slope cut of $\beta_{\rm UV}<-0.37$ (which correspond to the color cuts F277W-F444W > 1.0 mag and F150W-F200W < 0.5 for sources at $z\sim5$).  The bands used to measure the two slopes shift as a function of redshift as illustrated in Figure \ref{fig:QSO_SED_wFilters}, ensuring that we fit the same portion of the spectral energy distribution of all sources from $z=2$ out to $z\sim10$.  

 Figure \ref{fig:slopes} shows the resulting $\beta_{\rm UV}$ versus $\beta_{\rm opt}$ distribution for all galaxies at $z>2$ that are detected with a signal-to-noise ratio (SNR) greater than 12 in the F444W band.  
We find 791 
sources with $\beta_{\rm opt}>0$, $\beta_{\rm UV}<-0.37$, and SNR$_{\rm F444W}>12$ in the six fields we examined.

\begin{figure*}
\centering
\includegraphics[width=\linewidth]{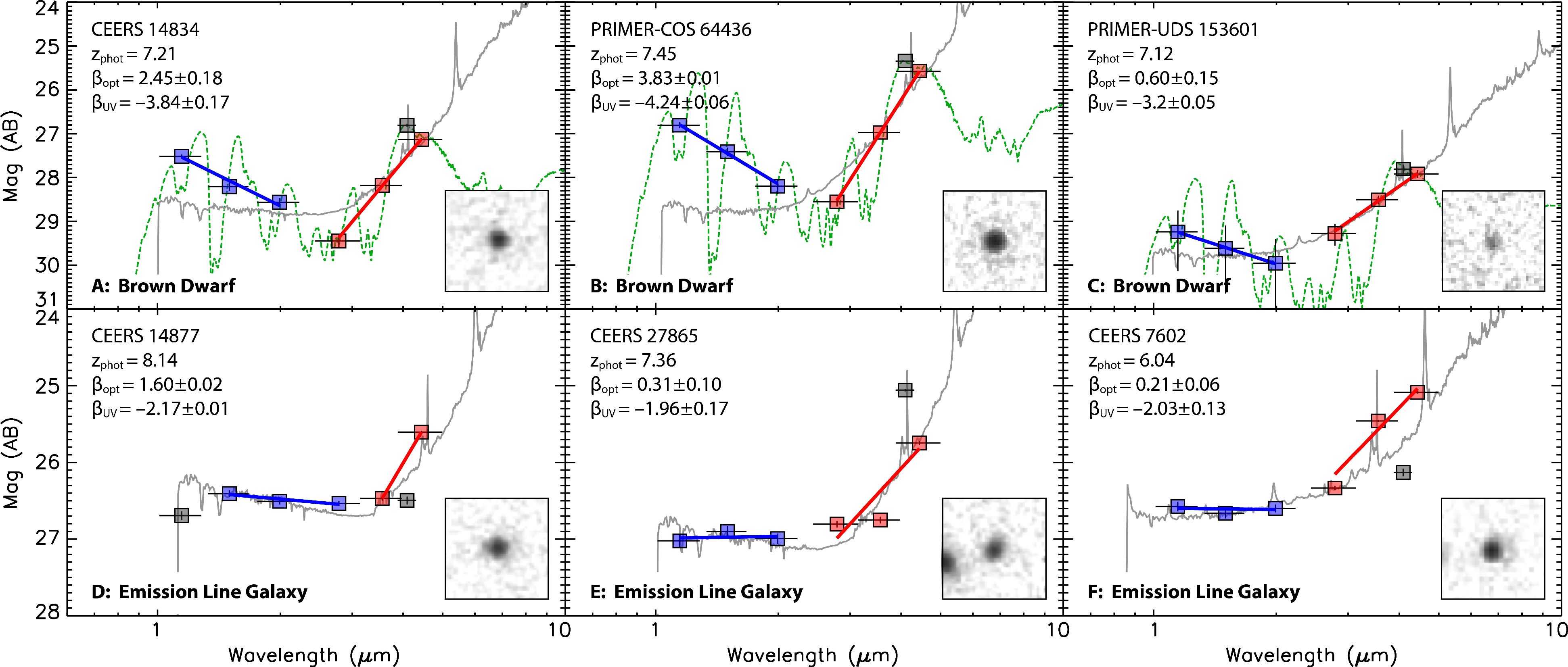}
\caption{Examples of our continuum slope fits for sources removed from our sample.  The bands used to measure the rest-frame UV and optical slopes are shown as blue and red squares, respectively.  Panels A, B and C show examples of candidate brown dwarfs cut from the sample due to their low rest-frame UV continuum slopes.  The dashed green line shows the best-fit LOWZ brown dwarf atmosphere model from \citet{Meisner21}. Panels D, E, and F show examples of sources whose rest-frame optical continuum slope may be boosted due to strong line emission in one or more bands. The F444W image cutouts of each source are $1^{\prime\prime}\times1^{\prime\prime}$ in size. The best-fit SEDs shown in light grey are galaxy plus QSO hybrid models (see \S3 for details). }
\label{fig:example_fits2}
\end{figure*}

\begin{figure*}
\centering
\includegraphics[width=\linewidth]{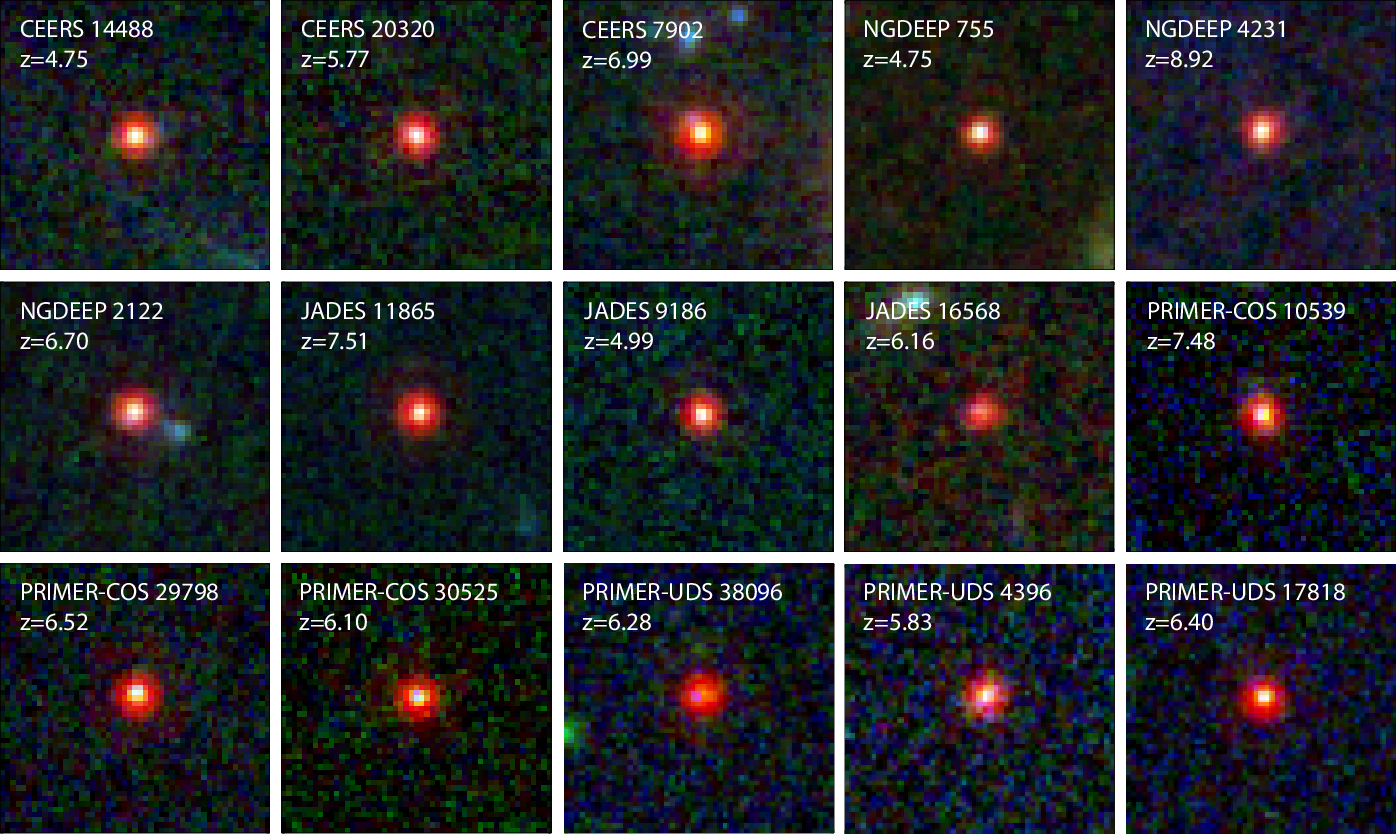}
\caption{Color image cutouts of a subset of our LRDs. The RGB images are composed of images in the F115W, F200W, and F444W filters. All images are $1\farcs5\times~1\farcs5$ in size.}
\label{fig:color_thumbs}
\end{figure*}

From this sample, we select sources with compact morphologies by applying an additional cut on half-light radius, $r_{h}$, as measured by SExtractor in the F444W band.  This cut is magnitude dependent and is designed to follow the tilt of the stellar locus apparent in the $m_{\rm F444W}$ versus $r_{h}$ parameter space, which can be seen in Figure \ref{fig:mag-size}.  To account for the increased scatter in size measurements at faint magnitudes, we select compact sources as those with $r_{h}$ within 1.5 times of the stellar locus (the dotted line in Figure \ref{fig:mag-size}).  
This size cut excludes 189 sources. This amounts to 24\% of the sample, which indicates that most objects selected to have red optical and blue UV colors also have compact morphologies.  This agrees with the findings of \citet{Barro23}, whose sample of EROs is largely unresolved.
Our size cut reduces our sample size from 791 to 602 sources.  

We apply two further selection criteria to remove potential contaminants.  The first is an additional cut on $\beta_{\rm UV}$ to remove brown dwarfs, which have similar near-infrared colors as reddened AGN, but appear significantly bluer at shorter wavelengths \citep{Langeroodi_2023}. 
To account for this, we exclude from our sample all sources with $\beta_{\rm UV} < -2.8$, which corresponds to the color cut ${\rm F115W-F200W} < -0.5$ employed by \citet{Greene_2023}.  Examples of potential brown dwarfs removed from our sample based on its $\beta_{\rm UV}$ slope are shown in panels A, B, and C of Figure \ref{fig:example_fits2}.  Each panel shows both our best-fit galaxy plus AGN hybrid model and the best-fit LOWZ brown dwarf atmosphere model from \citet{Meisner21}.
This color cut removes 63 objects and reduces the sample size to 539 sources.

The second refinement to our selection is meant to identify sources whose $\beta_{\rm opt}$ may be boosted due to strong line emission affecting one or more bands.  The SEDs of three such sources are shown in Figure \ref{fig:example_fits2}.  Panel D and E show sources with a flat spectral slope between F277W and F356W, followed by an increase in flux at F444W that can be explained by contamination from strong \OIII+\Hb~lines given the redshift of the two sources. The SED of the source in Panel F shows an elevated flux in F356W and F444W relative to F410M, which can be explained by \OIII+\Hb~and \Ha+\NII~emission boosting the flux in each band.  To remove such sources, we impose an additional requirement that $\beta_{\rm F277W-F356W} > -1$ and $\beta_{\rm F277W-F410M} > -1$ (when F410M photometry is available, which is true for all fields except NGDEEP).  For sources at $z>8$, only the second condition is imposed.  This requirement ensures that sources have a rising continuum over the entire wavelength range in which the optical continuum slope is measured.  The equivalent color cuts are ${\rm F277W-F356W} > 0.53$ and ${\rm F277W-F410M} > 0.84$.  This additional cut removes 198 sources, including all three sources shown in Figure \ref{fig:example_fits2}, resulting in a final sample size of 341 sources. 

In summary, our primary selection criteria are:
\begin{enumerate}[label=(\roman*),leftmargin=2cm]
    \item ${\rm SNR_{F444W}} > 12 \, $
    \item $\beta_{\rm opt}>0 \,$
    \item $-2.8 < \beta_{\rm UV}<-0.37 \,$
    \item $r_{h} < 1.5~r_{h,~{\rm stars}}.$
\end{enumerate}

To this, we add the following conditions aimed at removing sources whose optical continuum slope is likely boosted by strong line emission:
\begin{enumerate}[label=(\roman*),leftmargin=2cm]
\setcounter{enumi}{4}
\item $   \beta_{\rm F277W-F356W}>-1~~({\rm only~at}~z < 8), $
\item $   \beta_{\rm F277W-F410M}>-1~~({\rm when~F410M~available}). $
\end{enumerate}

\begin{deluxetable*}{lccccccccccc}[t]
\tablenum{3}
\tablecolumns{7}
\tablecaption{Properties of our primary sample of Little Red Dots \label{tbl:sample}}
\tablehead{
 \colhead{ID} & \colhead{RA} & \colhead{Dec} & \colhead{$z_{\rm best}$} &  \colhead{$z_{\rm flag}$} & \colhead{$\beta_{\rm UV}$} & \colhead{$\beta_{\rm opt}$} & \colhead{$m_{\rm F444W}$}& \colhead{$M_{\rm UV}$} & \\ 
 \colhead{} & \colhead{} & \colhead{(J2000)} & \colhead{(J2000)} & \colhead{} & \colhead{} & \colhead{} & \colhead{(AB mag)} & \colhead{(AB mag)}  }
\startdata
       CEERS 260 & 214.805768 &  52.878048 &  7.93 &  1 & $-2.23\pm0.18$ & $1.20\pm0.09$ & 26.13 & -19.57 \\
       CEERS 941 & 214.757973 &  52.839706 &  7.42 &  1 & $-1.68\pm0.18$ & $0.33\pm0.09$ & 25.69 & -19.70 \\
       CEERS 965 & 214.912514 &  52.949437 &  5.95 &  1 & $-1.87\pm0.47$ & $0.30\pm0.20$ & 26.86 & -18.11 \\
      CEERS 1669 & 214.817697 &  52.877855 &  5.47 &  1 & $-2.48\pm0.10$ & $0.04\pm0.06$ & 25.54 & -20.21 \\
      CEERS 1914 & 214.998408 &  53.004619 &  6.49 &  1 & $-2.05\pm0.24$ & $2.31\pm0.05$ & 24.83 & -19.02 \\
      CEERS 1954 & 215.002842 &  53.007588 &  7.48 &  1 & $-1.48\pm0.26$ & $0.86\pm0.11$ & 26.36 & -18.86 \\
      CEERS 2285 & 214.956832 &  52.973153 &  5.86 &  1 & $-1.61\pm0.26$ & $0.97\pm0.10$ & 25.80 & -18.64 \\
      CEERS 2520 & 214.844768 &  52.892101 &  8.74 &  1 & $-1.75\pm0.12$ & $2.18\pm0.19$ & 25.34 & -20.40 \\
      CEERS 3153 & 214.925762 &  52.945661 &  5.23 &  1 & $-1.63\pm0.59$ & $1.20\pm0.09$ & 25.99 & -17.24\\
      CEERS 4363 & 214.780370 &  52.834802 &  4.75 &  1 & $-1.93\pm1.39$ & $0.24\pm0.17$ & 26.73 & -17.01 \\
\enddata
\tablecomments{$z_{\rm flag}$: 1 = phot redshift, 2 = spect redshift. This table is available in its entirety at https://github.com/dalekocevski/Kocevski24 }
\end{deluxetable*}

\begin{figure}
\centering
\includegraphics[width=\linewidth]{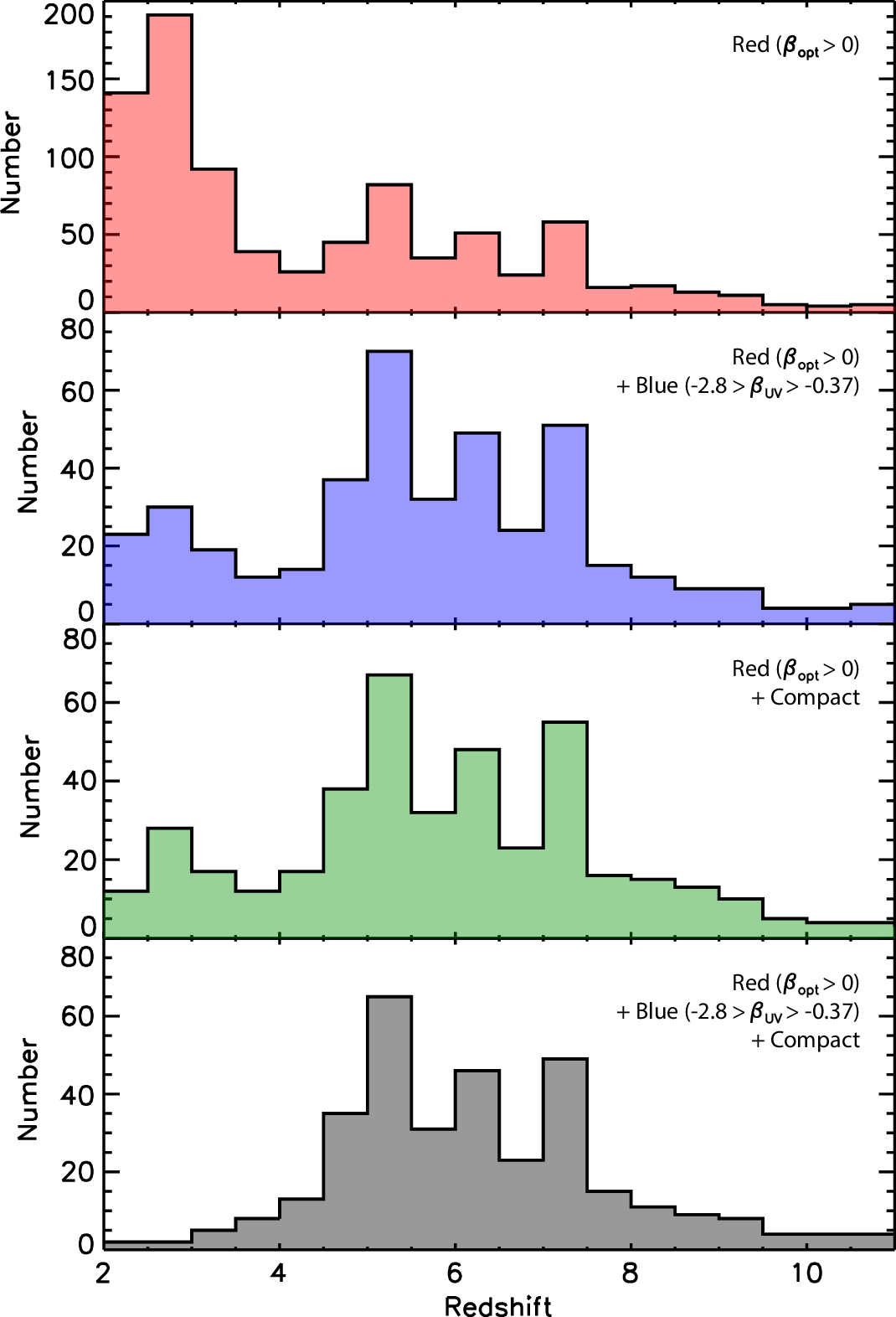}
\caption{The redshift distribution of sources with (\emph{top}) red rest-frame optical colors ($\beta_{\rm opt}>0$), (\emph{middle-top}) red rest-frame optical and blue rest-frame UV colors ($-2.8 < \beta_{\rm UV}<-0.37$), (\emph{middle-bottom}) red rest-frame optical color and a compact morphology, and (\emph{bottom}) red rest-frame optical and blue rest-frame UV colors and a compact morphology (see \S3).  The bottom panel is the redshift distribution of our final sample of LRDs.}
\label{fig:redshift_hist}
\end{figure}

\subsection{Vetting of $z>9$ Candidates} \label{sec:highz_selection}  

We apply additional scrutiny to sources in our sample with photometric redshifts of $z>9$ that is meant to remove potential lower-redshift interlopers \citep[e.g.,][]{Zavala23}.  Our selection criteria for such sources are similar to those of \cite{Finkelstein23} and are based on a combination of flux detection significances and the probability density function of the photometric redshift, $P(z)$.  First, we require a SNR $<3$ in all bands blueward of the Lyman break.  This includes F814W for $z>9$ and F115W for $z>11$.  Here the SNRs are measured in $0\farcs2$ diameter apertures in our non-PSF-matched images.  Second, the integrated probability that the source lies at $z>7$ must exceed 95\%, i.e. $\int P(z>7)>0.95$.  Third, the difference between the $\chi^{2}$ of the best-fit low-redshift ($z < 7$) and high-redshift models must exceed 4 ($\Delta \chi^{2} > 4$), corresponding to a $2\sigma$ significance \citep{Bowler20}.  Lastly, the (nonreduced) $\chi^2$ of the best-fit model must be $\chi^{2}<60$ to ensure a good fit to the photometry.

\subsection{SED Fitting}\label{sec:sed_fitting}

We model the SEDs of our sources using a custom $\chi^2$-minimization fit that employs multiple galaxy and AGN components in a manner similar to that described in \cite{Kocevski23b}.  We emphasize that these SED fits are not used for our sample selection and are instead only meant to visualize a possible SED consistent with all of our observed photometry in various figures throughout this paper. For the AGN component, we use the composite quasar spectrum of \cite{VB_2001} that is reddened by up to 4.5 magnitudes of visual extinction using a \cite{calzetti00} attenuation law.  To this, we add dust-free, scattered AGN emission that ranges from 0 to 10\% of pre-reddened AGN light.  For the stellar population, we assume a \citet{chabrier03} initial mass function, \citet{BC03} stellar population models, fixed solar metallicity, \citet{calzetti00} dust attenuation, and a delayed-$\tau$ star formation history with $\tau$ in the range of 0.1 Gyr to the age of the Universe at the source redshift (e.g., \citealt{maraston10}).

To determine various physical properties of the LRDs in our sample, including their host stellar masses, we model their SEDs using {\tt CIGALE v2022.1} \citep{Boquien19,Yang20,Yang22}.  For this modeling, we adopt parameters similar to those used in \cite{YangG23}.  We use the standard delayed-$\tau$ module {\tt sfhdelayed} for the star formation history, a \citet{chabrier03} initial mass function with a solar metallicity ($Z = 0.02$), and \citet{BC03} for the simple stellar population (SSP) module.  We add nebular emission using the {\tt nebular} model \citep{VillaV21} and use the {\tt dustatt\_modified\_starburst} module for the dust attenuation.  For the AGN component, we adopt the {\tt skirtor2016} module based on a clumpy torus model from \cite{Stalevski12,Stalevski16}. The relative strength between the AGN and galaxy components, set by the {\tt frac$_{\rm AGN}$} parameter, is allowed to vary from 0 to 0.99.

\section{Results}\label{sec:result}

Using the continuum slope and size criteria outlined in \S\ref{sec:sample}, we have identified 341 candidate obscured quasars with compact morphologies and red optical and blue UV colors.  These sources are split among the fields and datasets we examined as follows: 117 from PRIMER-UDS, 81 from PRIMER-COSMOS, 64 from CEERS, 46 from JADES, 23 from UNCOVER, and 10 from NGDEEP.  
The coordinates, redshifts, and best-fit continuum slopes are reported in Table \ref{tbl:sample}.  Color images of a subset of these sources, drawn from each dataset, are shown in Figure \ref{fig:color_thumbs}.  
In the following sections, we compare our sample to previous LRD compilations,  examine the redshift distribution and number density of our sample in greater detail, present a case study of the first X-ray detected LRD, and discuss broad emission line detections among our sample.

\subsection{Comparison to Previous Samples} \label{sec:sample_comp}  

In Figures \ref{fig:slopes} and \ref{fig:mag-size}, we highlight the sample of LRDs presented in three previous studies: \citet[hereafter B23]{Barro23} in the EGS field selected using CEERS data, \citet[hereafter PG24]{Perez-Gonzalez24} in the GOODS-S field selected using JADES data, and \citet[hereafter L23]{Labbe23} in the Abell 2744 field using data from the UNCOVER survey.  The B23 sample consists of 37 EROs selected to have NIRCam colors of ${\rm F277W}-{\rm F444W}>1.5$ and ${\rm F444W}<28$.  
We recover all 37 of these sources, however two (B23 ID 48777 and 72897) are excluded from our primary sample as potential brown dwarfs since they have UV spectral slopes of $\beta_{\rm UV}<-2.8$.   Of the remaining 35 sources, all but four have $\beta_{\rm opt}>1$, making them among the reddest in our CEERS sample.

The PG24 sample consists of 31 sources selected to have NIRCam colors of ${\rm F277W}-{\rm F444W}>1.0$, ${\rm F150W}-{\rm F200W}<0.5$ and ${\rm F444W}<28$.  Our primary sample includes 21 of these sources.  Two were excluded for failing our SNR cut (i.e., ${\rm SNR_{F444W}}> 12$), three were excluded for having $\beta_{\rm opt}<0$, and one source was excluded due to having a photometric redshift of $z<2$. Of the remaining 25 sources, four were removed as their optical continuum slopes are potentially contaminated by strong line emission.  

The L23 sample consists of 33 sources selected to have one of two possible red optical and blue UV color combinations (referred to as \emph{red1} and \emph{red2}), as well as a compact morphology.  The \emph{red1} criteria include ${\rm F115W}-{\rm F150W}<0.8$ \& ${\rm F200W}-{\rm F277W}>0.7$ \& ${\rm F200W}-{\rm F356W}>1.0$, while the \emph{red2} criteria are ${\rm F150W}-{\rm F200W}<0.8$ \& ${\rm F277W}-{\rm F356W}>0.7$ \& ${\rm F277W}-{\rm F444W}>1.0$.   These color criteria are the same as those used in \cite{Kokorev24}.

Of these sources, two were excluded from our sample as they were spectroscopically confirmed to be brown dwarfs by \citet[L23 IDs 23778 and 29466]{Greene_2023} and another two were flagged as potential strong emission line galaxies (L23 IDs 8602 and 15798).  Six sources were excluded for failing our optical slope cut (L23 IDs 2476, 6151, 37108, 46991, 49555 and 49567).  

These six sources highlight a notable difference between our selection criteria and that of L23 and \cite{Kokorev24}.  The SEDs of these sources exhibit a jump between F200W and F277W, consistent with the Balmer break falling between the two bands, but otherwise have flat SEDs blueward and redward of the break.  Such SEDs will be selected by the two red color criteria in \emph{red1}, namely ${\rm F200W}-{\rm F277W}>0.7$ \& ${\rm F200W}-{\rm F356W}>1.0$.  However, when the continuum is fit using three bands (i.e., F200W, F277W, and F356W), the flat SED after the break reduces the inferred spectral slope and ultimately drops it below our selection threshold.  Overall, 18\% (6/33) of the L23 sample failed to make our selection for this reason.  Our continuum slope fitting technique appears less susceptible to contamination from galaxies with strong breaks, especially at $z\sim5-6$, where the Balmer break falls between the F200W and F277W bands.  The SEDs of such sources are likely galaxy-dominated and not the AGN candidates with rising red continua that our LRD selection is intended to identify.

\begin{figure*}
\centering
\includegraphics[width=\linewidth]{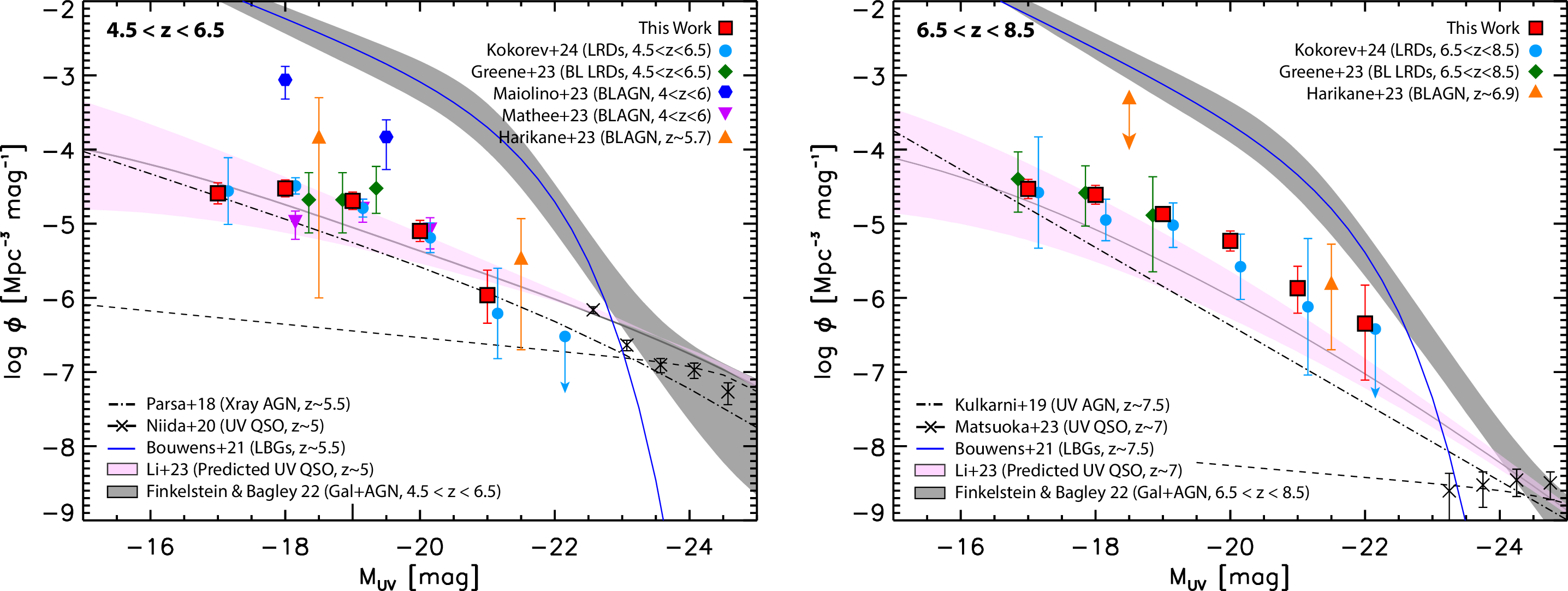}
\caption{The UV luminosity function of our sample measured at rest-frame 1450\AA~in two redshift bins: $4.5<z<6.5$ (\emph{left}) and $6.5<z<8.5$ (\emph{right}). We find good agreement with previous photometric and spectroscopic compilations of LRDs \citep{Matthee_2023, Greene_2023, Kokorev24}.  We find LRDs are $\sim4$ and $\sim10$ times more numerous at $M_{\rm UV}=-19$ than X-ray AGN at $z\sim5$ \citep{Parsa18} and UV-selected AGN at $z\sim7$ \citep{Kulkarni19}, respectively.  We also note fair agreement with the model predictions of \cite{Li_Inayoshi23}.}

\label{fig:LF}
\end{figure*}

\subsection{Redshift Distribution} \label{sec:redshift_dist}  

Our spectral slope fitting technique, which uses shifting bandpasses to sample the same rest-frame emission blueward and redward of 3645\AA, allows us to self-consistently search for AGN candidates with red optical and blue UV colors over a wide range of redshifts.  The bottom panel of Figure \ref{fig:redshift_hist} shows the redshift distribution of our resulting sample.  We find sources with this unique spectral shape and compact morphology emerge at $z\sim8$ and have a median redshift of $z\sim6.4$.  Their number density then undergoes a rapid decline at $z\sim4$. 

Our methodology of using shifting bandpasses was meant to help identify lower-redshift analogues of the LRDs previously identified in the literature.  However, we identify very few unresolved sources with red optical and blue UV colors at $z<4$: only 17 are found in our sample.  This is not due to sources becoming more extended at lower redshifts.  Even without our size cut, we find few sources with red optical and blue UV colors at the lowest redshifts we examined.  This fact can be seen in the top and middle-top panels of Figure \ref{fig:redshift_hist}, which show the redshift distribution of all red sources ($\beta_{\rm opt}>0$) and red plus blue sources ($-2.8 < \beta_{\rm UV}<-0.37$) regardless of their effective radii.  While we find a growing number of optically red sources at $z<4$, most ($82\%$) are removed from our sample based on our UV color cut.  Only $14\%$ of the optically red and UV blue sources are cut based on their size.  
Our findings suggest the population of LRDs with a UV-excess emerge in large numbers for the first time at about $z\sim4$.  This may explain why such sources were not previously identified using imaging from HST and Spitzer/IRAC.  

At the high-redshift end, our selection becomes increasingly incomplete as long-wavelength filters start to shift blueward of the Balmer break.
For example, the F356W band probes fully blueward of 4000\AA~at $z\sim9$.  This will flatten the inferred spectral slope (i.e., lower $\beta_{\rm opt}$) for sources that are blue in the rest-frame UV and increase the likelihood that sources will fall below our continuum slope cut.  This may help explain the rapid decline in the redshift distribution above $z\sim8$.

\subsection{Number Density} \label{sec:NumDensity}  

\begin{table}
\renewcommand\thetable{4} 
\caption{Rest-frame UV luminosity function of our sample of LRDs in the redshift range $4.5<z<6.5$ and $6.5<z<8.5$. }\vspace{0mm}
\begin{center}
\vspace{-0.1in}
\begin{tabular}{ccc}
\hline
\hline
$M_{\rm UV}$ & $N$ & $\Phi$      \\
  (AB Mag)   &   &  (Mpc$^{-3}$ mag$^{-1}$)  \\
\hline 
\multicolumn{3}{c}{$4.5<z<6.5$} \\
\hline
-17.0 &  22 &  $-4.59\pm 0.10$  \\ 
-18.0 &  62 &  $-4.53\pm 0.06$  \\ 
-19.0 &  56 &  $-4.69\pm 0.06$ \\
-20.0 &  22 &  $-5.10\pm 0.10$ \\ 
-21.0 &   3 &  $-5.96^{+0.29}_{-0.34}$ \\ 
\hline 
\multicolumn{3}{c}{$6.5<z<8.5$} \\
\hline
-17.0 &  14 &  $-4.53\pm 0.13$  \\ 
-18.0 &  33 &  $-4.61\pm 0.08$ \\ 
-19.0 &  28 &  $-4.87^{+0.09}_{-0.08}$ \\
-20.0 &  13 &  $-5.23^{+0.13}_{-0.14}$ \\ 
-21.0 &   3 &  $-5.87^{+0.29}_{-0.34}$ \\ 
-22.0 &   1 &  $-6.35^{+0.52}_{-0.76}$ \\ 
\hline
\end{tabular}
\label{tab:lumfunc}
\end{center}
\end{table}

In this section, we present the rest-frame UV luminosity function (LF) of the LRDs in our sample. To determine the rest-UV magnitude of our sources, we extrapolate our best-fit to the blue continuum to rest-frame 1450\AA.  The resulting absolute magnitudes at this wavelength are listed in Table \ref{tbl:sample}.  To compute the number density of our sample, we employ the $1/V_{\rm max}$ method \citep{Schmidt68}:

\begin{equation}
\Phi(m)~{\rm d}m = \sum\limits_{i} \frac{1}{V_{i}(M_{i})}
\end{equation}

\noindent where $V_{i}(M_{i})$ is the maximum comoving volume in which source $i$ with absolute magnitude $M_{i}$ is detectable.  For a given survey area $\Delta \Omega$ and a redshift range $z_{\rm min} < z < z_{\rm max}$, the comoving volume is computed as:

\begin{equation}
V_{i} = \frac{c~\Delta \Omega_{i}}{H_{0}}\int_{z_{\rm min}}^{z_{{\rm max}, i}}\frac{D_{L}(z)^{2}}{(1+z)^{2}}[\Omega_{m}(1+z)^{3} +  \Omega_{\Lambda}]^{-1/2}.
\end{equation}

For this calculation, we exclude sources selected from the UNCOVER dataset due to the variable magnification in the Abell 2744 field.  We account for magnitude incompleteness effects by computing the total survey area ($\Delta \Omega_{i}$) and maximum redshift ($z_{{\rm max},i}$) over which each source could have been detected.  Given that our sources are substantially fainter in the rest-UV versus the rest-optical, the latter is largely driven by the requirement that sources have rest-UV flux measurements with which to measure a continuum slope. 

We calculate the LF using a bin size of 1.0 magnitude and over two redshift ranges: $4.5 < z < 6.5$ and $6.5 < z < 8.5$.  The uncertainty on the resulting number densities is estimated with Poisson statistics corrected for low number counts following \cite{Gehrels86}.  Our calculated number densities are listed in Table \ref{tab:lumfunc} and shown in Figure \ref{fig:LF}. 

We find good agreement with the number densities reported by \cite{Kokorev24} 
for their color-selected samples of LRDs, as well as those inferred by \cite{Matthee_2023} and \cite{Greene_2023} for their spectroscopically-confirmed samples of dust-reddened, broad-line AGN at both $z\sim5$ and $z\sim7$.  Comparing to the results \citet{Harikane23} and \citet{Maiolino23}, we find that if LRDs are indeed powered by reddened AGN, they would make up 10-15\% of the overall faint, broad-line AGN population detected by JWST.

\begin{figure*}
\centering
\includegraphics[width=\linewidth]{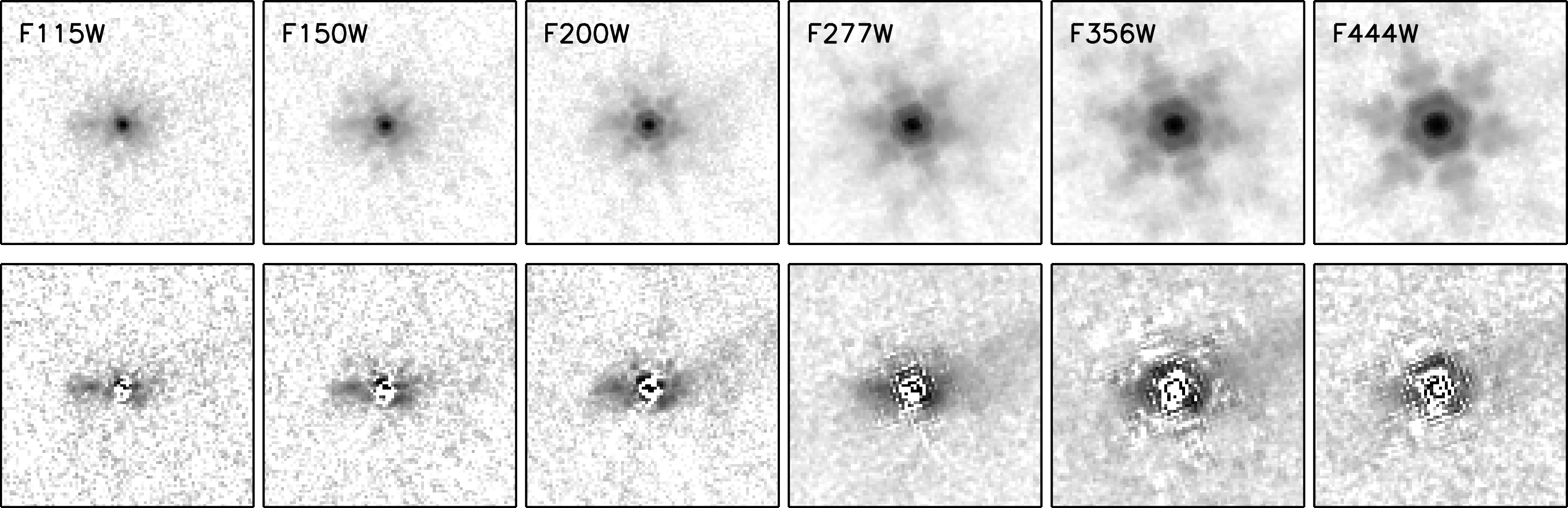}
\caption{(\emph{top}) Multi-wavelength image cutouts of the source PRIMER-COS 3866 (CXOC100024.2+022510).  The source is one of only two little red dots in our sample which are X-ray detected.  All images are $2\farcs5\times2\farcs5$ in size. (\emph{bottom}) Images of the underlying host galaxy after subtracting our best-fit point-source model.}
\label{fig:galfit_montage}
\end{figure*}

\begin{figure}
\centering
\includegraphics[width=\linewidth]{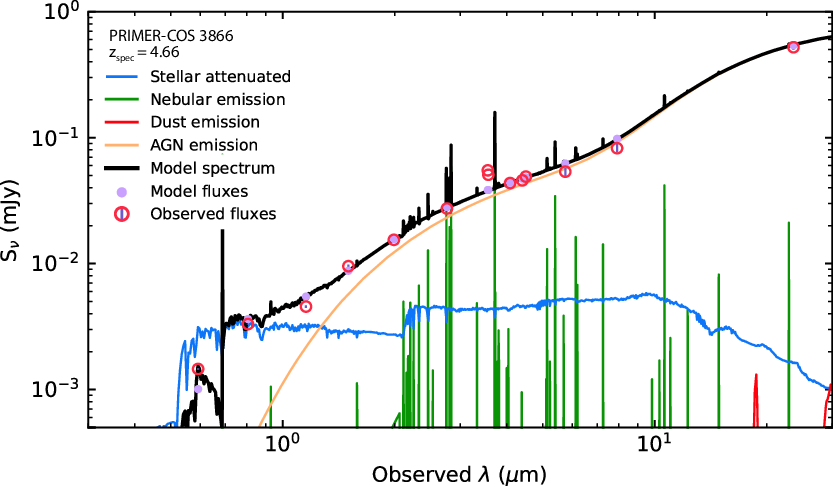}
\caption{Best-fit galaxy plus AGN SED model from {\tt CIGALE} of the X-ray detected, obscured AGN PRIMER-COS 3866.  The red and purple points indicate the observed and model flux densities, respectively. The black curves represent the total model SEDs. The orange and blue curves indicate the AGN and attenuated stellar components, respectively.}
\label{fig:cigale_fit}
\end{figure}

\begin{figure}
\centering
\includegraphics[width=\linewidth]{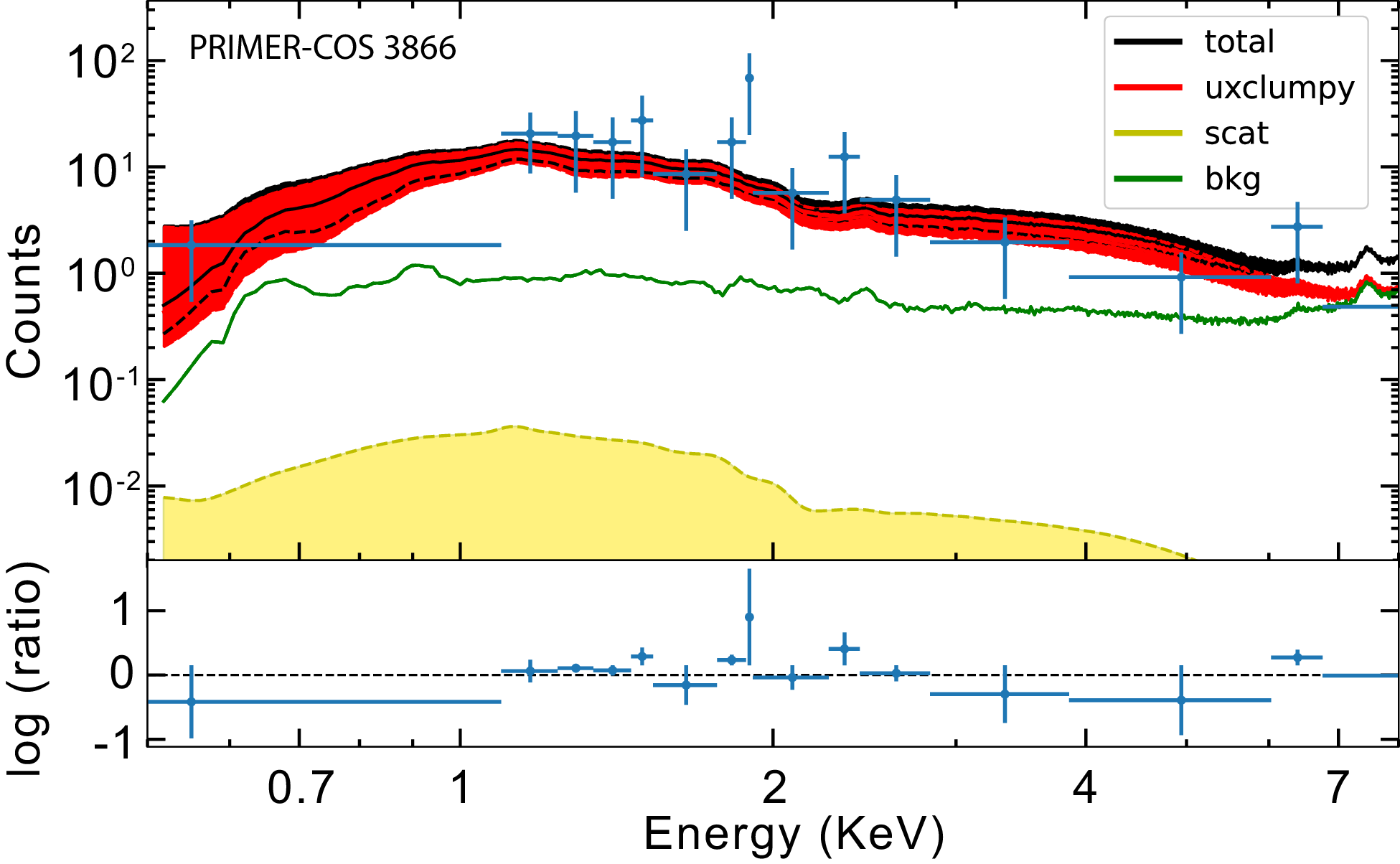}
\caption{Results of the X-ray spectral analysis of PRIMER-COS 3866 by \cite{Laloux23} using BXA with the UXCLUMPY model.  The blue crosses are the extracted X-ray spectrum grouped to yield an SNR above 1 per bin. The red line corresponds to the UXCLUMPY model, the yellow line represents the soft scattering, and the green line is the background model. The sum of all three components above is shown with the black line. The shaded regions correspond to the $1\sigma$ confidence interval of the corresponding component. The lower panel plots the logarithmic ratio between the X-ray spectrum and the best-fitting model as a function of the energy.}
\label{fig:xray_spectrum}
\end{figure}

In agreement with previous results \citep{Kocevski23b, Matthee_2023,Greene_2023}, we find that our candidate faint, red AGN are 2-3 dex more numerous than would be expected based on simple extrapolations of the quasar luminosity function measured from the ground at $z\sim5-7$ \citep{Niida20, Matsuoka23}. However, we find better agreement with the number densities measured for lower-luminosity AGN at these redshifts.  The number density of LRDs at $z\sim5$ is $4\times$ higher than that of X-ray AGN reported in \cite{Parsa18}.  This offset becomes more pronounced at $z\sim7$, where we infer LRDs to be $10\times$ more common than color-selected AGN at this redshift \citep{Kulkarni19}.  We note that we do find fair agreement with the predicted quasar luminosity function at the faint regime by \cite{Li_Inayoshi23}, which is the result of a semianalytical model for BH formation and growth that considers multiple accretion bursts with variable Eddington ratios and is primarily calibrated with the quasar abundance at the bright end of $M_{\rm UV}<-24$.  Finally, At $z\sim7$, we find that our sources constitute $\sim3\%$ of the overall galaxy population at $M_{\rm UV}=-20$.  This fraction rises to $\sim10\%$ at $M_{\rm UV}=-22$, although this is based on a single bright source.

\subsection{X-ray Detected Little Red Dots} \label{sec:xray_example} 

We cross-matched our sample of 341 candidate red AGN with the publicly available X-ray source catalogs from the Chandra Deep Field South (CDFS) observations in GOODS-South \citep{Luo17}, the AEGIS-XD survey \citep{nandra15}, the X-UDS survey \citep{kocevski18}, and the C-COSMOS survey \citep{Elvis09}, which have characteristic exposures times of 7 Msec, 800 ksec, 600 ksec, and 200 ksec, respectively.  All of the 341 LRDs in our primary sample fall within the footprint of these X-ray observations.  We find only two objects that are directly detected: PRIMER-COS 3866 and JADES 21925.  These sources have redshifts of $z=4.66$ and $z=3.1$, respectively and appear to be bright, lower-redshift analogs of the LRDs identified at $z=5-7$.  As such, they may provide unique insight into the nature of these objects.  In this section, we examine the properties of both sources in greater detail.

\subsubsection{PRIMER-COS 3866 at $z=4.66$}

PRIMER-COS 3866 was first identified as an X-ray emitter by \citet{Elvis09} as CXOC100024.2+022510 and later spectroscopically confirmed to be at $z=4.66$ by \cite{Civano11}. The source has a F444W magnitude of 19.75, making it the brightest object in our sample.  Images of PRIMER-COS 3866 in multiple NIRCam bands are shown in Figure \ref{fig:galfit_montage}.  In addition to its strong NIRCam detection, the source is also detected in the rest-frame UV with \emph{Hubble} ACS imaging and in the near- and mid-infrared with \emph{Spitzer} IRAC and MIPS imaging.  

\begin{figure*}
\centering
\includegraphics[width=\linewidth]{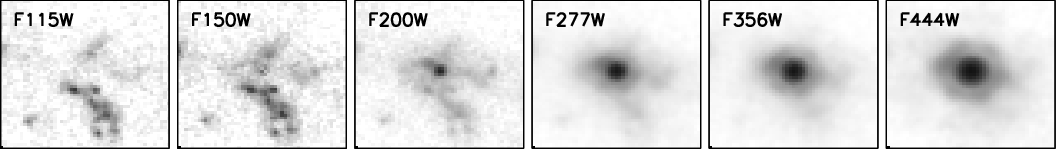}
\caption{Multi-wavelength image cutouts of the source JADES 21925.  The source is one of only two little red dots in our sample which are X-ray detected.  All images are $1\farcs5\times1\farcs5$ in size.}
\label{fig:JADES21925_images}
\end{figure*}

\begin{figure}
\centering
\includegraphics[width=\linewidth]{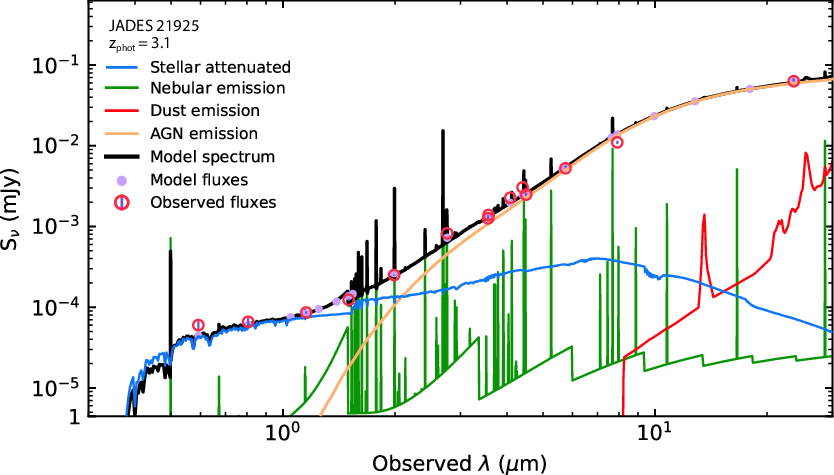}
\caption{Best-fit galaxy plus AGN SED model from {\tt CIGALE} of the X-ray detected, obscured AGN JADES 21925.  The red and purple points indicate the observed and model flux densities, respectively. The black curves represent the total model SEDs. The orange and blue curves indicate the AGN and attenuated stellar components, respectively.}
\label{fig:JADES21925_SED}
\end{figure}

In Figure \ref{fig:cigale_fit}, we show the best-fit galaxy plus AGN SED model produced by fitting the available photometry with {\tt CIGALE v2022.1} using the parameters listed in \S3.3.  For this fit, we use \emph{Spitzer} IRAC and MIPS fluxes from the ‘super-deblended’ infrared photometric catalogue of \cite{Jin18}.
We find the $3-30\mu m$ emission of the source is well-fit, and dominated by, AGN emission that is moderately reddened by polar dust with an $E(B-V)=0.40\pm0.01$.  The {\tt skirtor2016} model used for the AGN component is based on a clumpy torus model, which would suggest the observed extinction is due to torus-level obscuration as opposed to galaxy-wide obscuration.  The CIGALE results suggest the relatively blue color of the source in the rest-frame UV is due to emission from the host galaxy at $<1\mu m$. CIGALE places the stellar mass of the underlying galaxy at ${\rm log}(M_*/{\rm M}_{\odot})=10.6^{+0.2}_{-0.4}$.

Because there has been considerable debate as to the origin of the blue rest-frame UV emission from LRDs (e.g.~\citealt{Kocevski23b}), we used the {\tt GALFIT} software \citep{peng02} to subtract off emission from the central point source in several NIRCam bands to search for signs of the underlying galaxy.  For this modeling, we provide {\tt GALFIT} with empirical PSFs constructed from the PRIMER-COS mosaic and noise images that account for both the intrinsic image noise (e.g., background and readout noise) and added Poisson noise due to the objects themselves. The results of the point source subtraction can be seen in the lower panels of Figure \ref{fig:galfit_montage}.  We see clear signs of an elongated, potentially clumpy host galaxy emerging in the short-wavelength NIRCam images, in general agreement with our best-fit SED model.

Lastly, the X-ray emission from PRIMER-COS 3866 affords us an independent way to assess the obscured nature of the source.  The shape of an AGN’s X-ray spectrum can both reveal the presence of gas obscuring the central engine and provide a measure of its column density, $N_{\rm H}$.  An X-ray spectral analysis of PRIMER-COS 3866 was recently carried out by \cite{Laloux23}, where the UXCLUMPY clumpy torus model of \cite{Buchner19} was fit to the observed X-ray spectrum using the Bayesian X-ray Analysis (BXA) package \citep{Buchner14}.  The results of this analysis are shown in Figure \ref{fig:xray_spectrum}.  
The best-fit column density is found to be $\log\,(N_{\rm H}/{\rm cm}^{-2})=23.3^{+0.4}_{-1.3}$ , resulting in an obscuration-corrected X-ray luminosity of $\log\,(L_{2-10~{\rm keV}}/{\rm erg~s}^{-1}) = 44.7\pm0.2$.
The ratio of our measured extinction to our best-fit gas column density, $E(B-V)/N_{\rm H}$, is a factor of $\sim80$ below the Galactic standard value of $1.7\times 10^{-22}$ mag cm$^{-2}$ \citep{Savage79}, but in good agreement with the reduced $E(B-V)/N_{\rm H}$ ratios reported by \cite{Maiolino21} for various classes of luminous ($L_{\rm 2-10~keV} > 10^{43}$ erg s$^{-1}$) AGN.

These results confirm that PRIMER-COS 3866 harbors a luminous, moderately obscured AGN, in agreement with our SED modeling of the source.
The best-fit $2-10$ keV luminosity, combined with a bolometric correction $L_{\rm bol} / L_{\rm 2-10 kev} \sim 20$ \citep{Duras20} appropriate for this luminosity class, leads to an estimate of the black hole mass of $\sim 8 \times 10^7 \, {\rm M}_\odot$, assuming it is accreting at the Eddington limit.  Given the estimated stellar mass of the host, this black hole mass places this source near the standard local relation between the stellar mass and black hole mass (see \S4.5.2 for additional details).

\subsubsection{JADES 21925 at $z=3.1$}
JADES 21925 was first identified as an X-ray emitter by \citet{Giacconi02} as CXOCDFSJ033220.9-275223 in the 1 Msec Chandra Deep Field South observations and more recently detected in the 7 Ms observations of the field presented in \citet{Luo17}.  Images of JADES 21925 in multiple NIRCam bands are shown in Figure \ref{fig:JADES21925_images}.  The source shows clear signs of an extended, blue structure surrounding a reddened point source that only becomes prominent at wavelengths longer than 2$\mu$m.  The photometric redshift of the source was reported as $z=1.5$ in the \citet{Luo17} catalog, which we revise upward to $z=3.1$ with the inclusion of JWST photometry from JADES.  Based on its X-ray detection, JADES 21925 is the lowest redshift LRD confirmed to host an AGN (a record it holds with RUBIES-BLAGN-1 \citep{Wang24}, which is found to have the same redshift; see \S4.5 below).

In Figure \ref{fig:JADES21925_SED}, we show the best-fit galaxy plus AGN SED model for JADES 21925 produced with {\tt CIGALE v2022.1} using the parameters listed in \S3.3.  For this fit, we use \emph{Spitzer} IRAC and MIPS fluxes from the photometric catalogue of \cite{Guo13}.  Like PRIMER-COS 3866, we find the $3-30\mu m$ emission of JADES 21925 is dominated by AGN emission, while the rest-frame UV emission is attributed to the host galaxy.  We find the extinction of the AGN emission by polar dust to be $E(B-V)=1.39\pm0.5$ and CIGALE's estimate for the stellar mass of the underlying galaxy is ${\rm log}(M_*/{\rm M}_{\odot})=8.84^{+0.15}_{-0.38}$.

\begin{figure}
\centering
\includegraphics[width=\linewidth]
{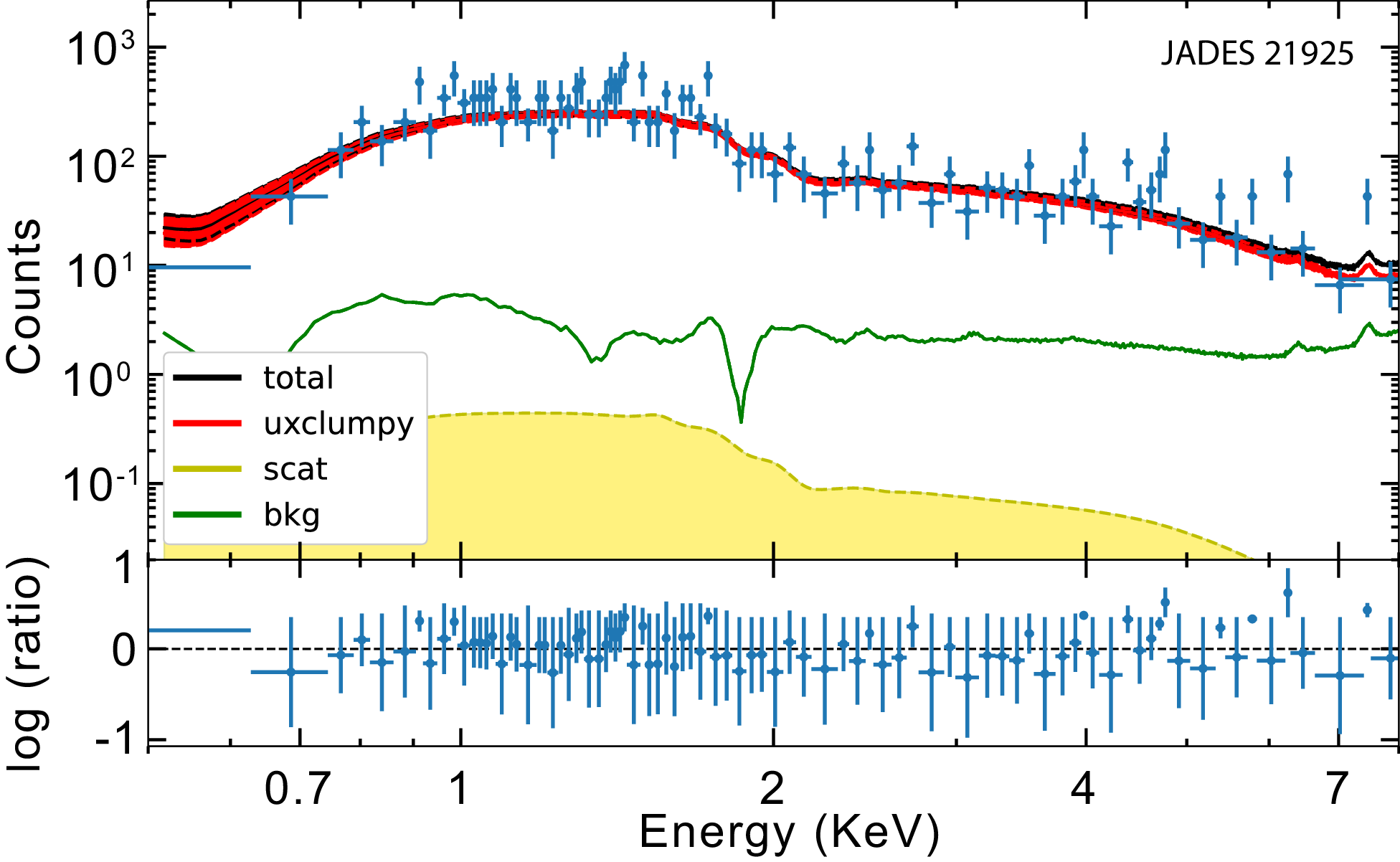}
\caption{Results of our X-ray spectral analysis of JADES 21925 using BXA with the UXCLUMPY model.  The blue crosses are the extracted X-ray spectrum grouped to yield an SNR above 1 per bin. The red line corresponds to the UXCLUMPY model, the yellow line represents the soft scattering, and the green line is the background model. The sum of all three components above is shown with the black line. The shaded regions correspond to the $1\sigma$ confidence interval of the corresponding component. The lower panel plots the logarithmic ratio between the X-ray spectrum and the best-fitting model as a function of the energy.}
\label{fig:xray_spectrum}
\end{figure}

\begin{figure}[t]
\centering
\includegraphics[width=\linewidth]{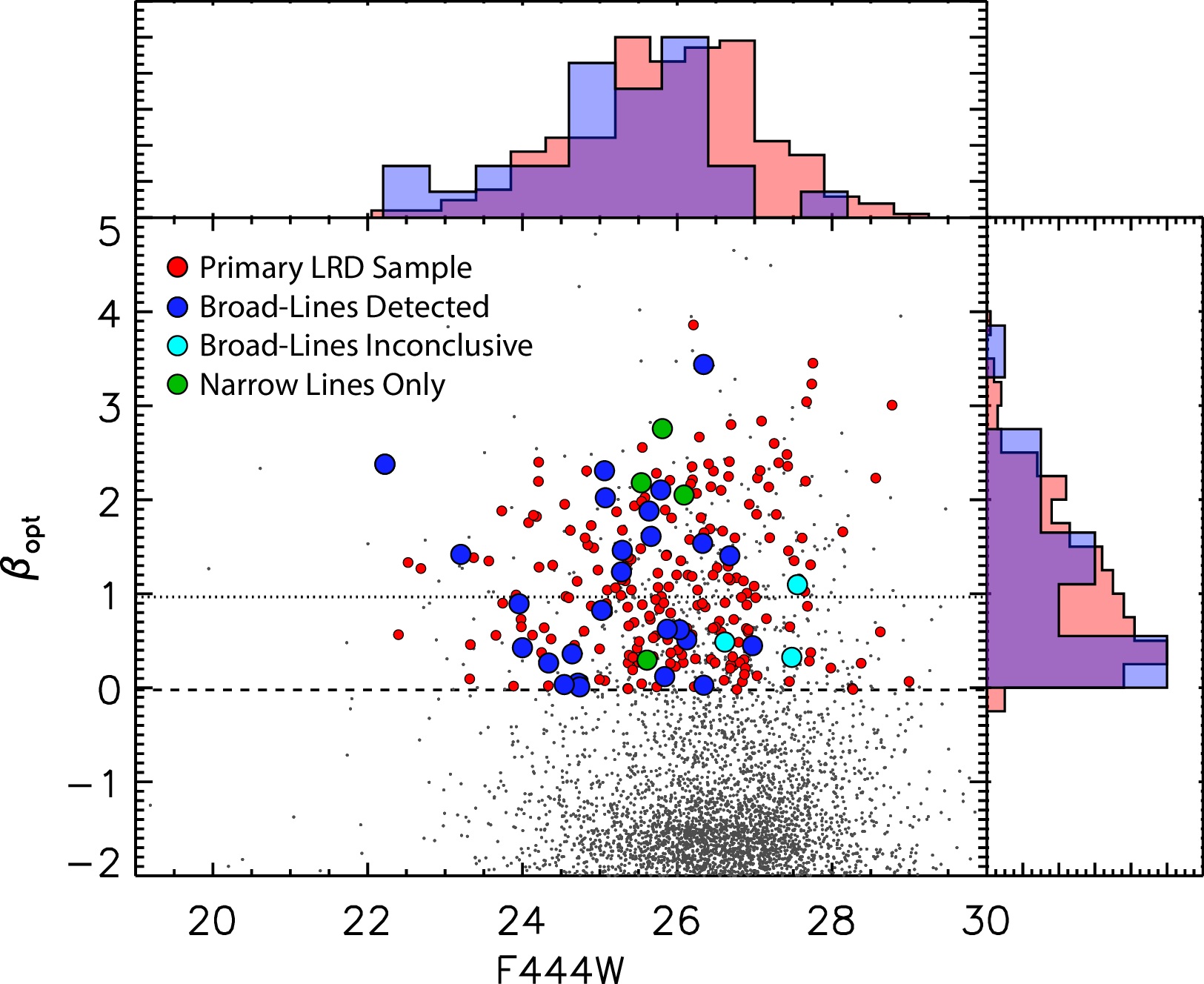}
\caption{Distribution of F444W magnitudes and $\beta_{\rm opt}$ slopes for our full LRD parent sample (red) and those with confirmed broad-line detections (blue).  We find that the broad-line LRDs span the full range of $\beta_{\rm opt}$ slopes found in our parent sample and are not limited to the reddest sources.}
\label{fig:mag-slope-blagn}
\end{figure}

We also performed an X-ray spectral analysis of JADES 21925 using the BXA package and the UXCLUMPY torus model of \cite{Buchner19}.  The model assumes a powerlaw with a photon index prior of $1.95\pm0.15$ reprocessed with a torus (including absorption, Compton scattering, and Fe fluorescence) and an additionally absorbed reflection component. The results of this analysis indicate JADES 21925 is moderately attenuated, with a best-fit column density of $\log\,(N_{\rm H}/{\rm \,cm}^{-2})=22.72^{+0.13}_{-16}$, resulting in an obscuration-corrected X-ray luminosity of $\log\,(L_{2-10~{\rm keV}}/{\rm \,erg\,s}^{-1}) = 43.73\pm0.06$.  The ratio of our measured extinction and column density again favors a $E(B-V)/N_{\rm H}$ ratio below that of the Galactic standard value.  Using the median of the ratios reported in \cite{Maiolino21} and our measured $E(B-V)$ extinction would predict a column density of $\log\,(N_{\rm H}/{\rm cm}^{-2}) = 22.5$, which agrees with our measured value, within the errors, better than the 21.6 cm$^{-2}$ obtained using the Galactic standard ratio.

Our measured X-ray luminosity, coupled with the relatively low stellar mass that we infer for the host from our SED fit, indicates that JADES 21925, like PRIMER-COS 3866, may harbor an overmassive black hole.  Using a bolometric correction of $L_{\rm bol} / L_{\rm 2-10 kev} \sim 20$ and assuming the system is accreting at its Eddington limit, results in a black hole mass of $\sim 8.5 \times 10^6 \, M_\odot$.  This would imply this source lies significantly above the local black hole versus galaxy stellar mass relationship, as seen in dozens of other $z>4$ galaxies discovered by JWST (see, e.g., \citealt{Pacucci_2023_overmassive}).
.

\subsubsection{Implications of X-ray Non-Detections}

We find that with the exception PRIMER-COS 3866 and JADES 21925, the vast majority of the LRDs in our sample are X-ray undetected.  This includes 55 LRDs from JADES and NGDEEP that are undetected in the deepest X-ray observations available, the 7 Msec CDFS observations.  Stacking the X-ray data at the location of these sources using CSTACK version 4.5 \citep{Miyaji_Griffiths08} results in a non-detection in both the soft (0.5-2 keV) and hard (2-8 keV) bands, in agreement with the results of several such stacking analyses reported in the literature \citep{Yue24, Ananna24, Maiolino24}.  This raises the question of why PRIMER-COS 3866 and JADES 21925 are the lone LRDs directly detected.  

\begin{figure}[t]
\centering
\includegraphics[width=\linewidth]{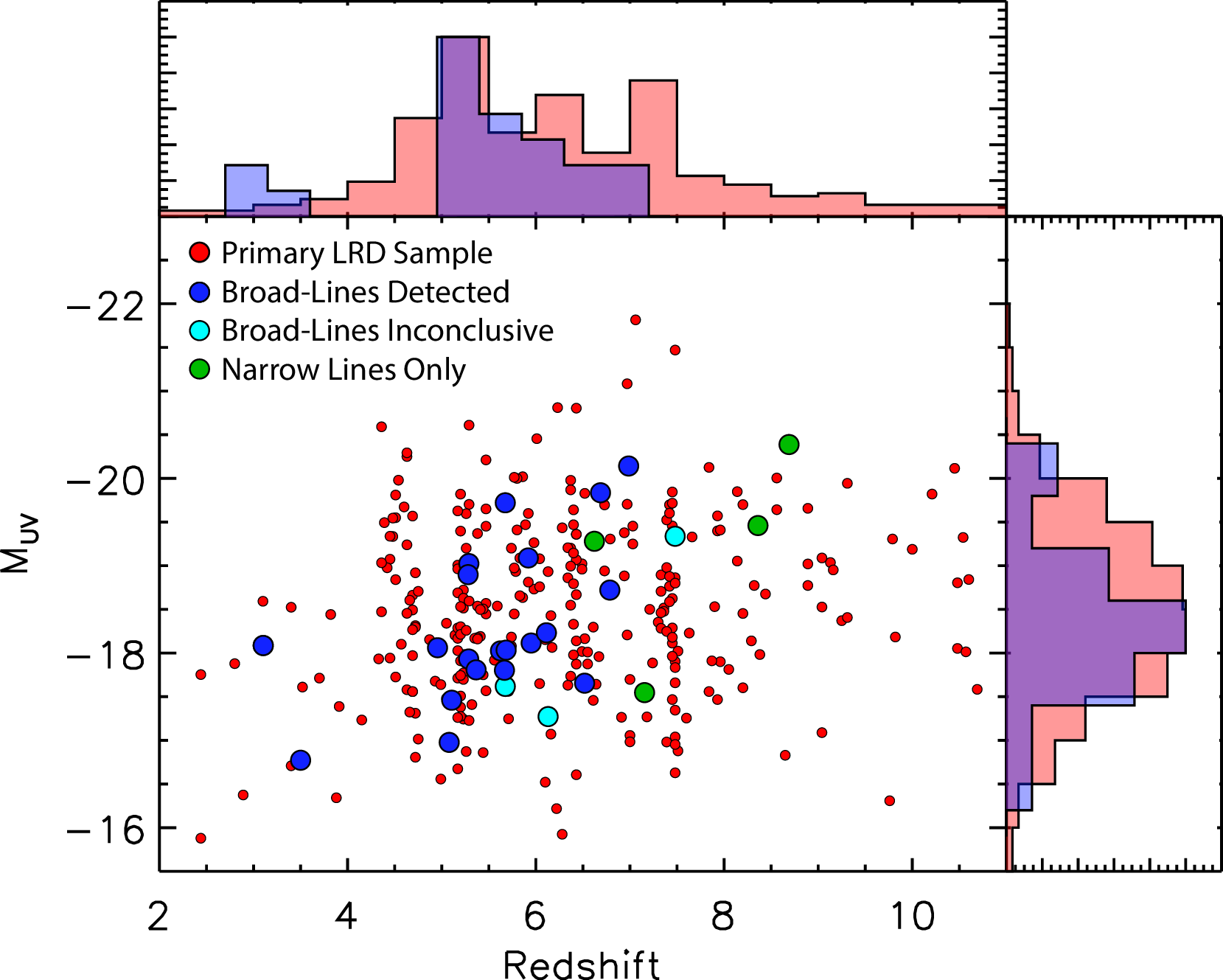}
\caption{Distribution of redshifts and $M_{\rm UV}$ magnitudes for our full LRD parent sample (red) and those with confirmed broad-line detections (blue).  We find that the broad-line LRDs  are skewed toward brighter sources and lower redshifts ($z<7$).  See \S\ref{sec:xray_example} for additional details.}
\label{fig:redshift-MUV-blagn}
\end{figure}

\begin{figure*}
\centering
\includegraphics[width=\linewidth]{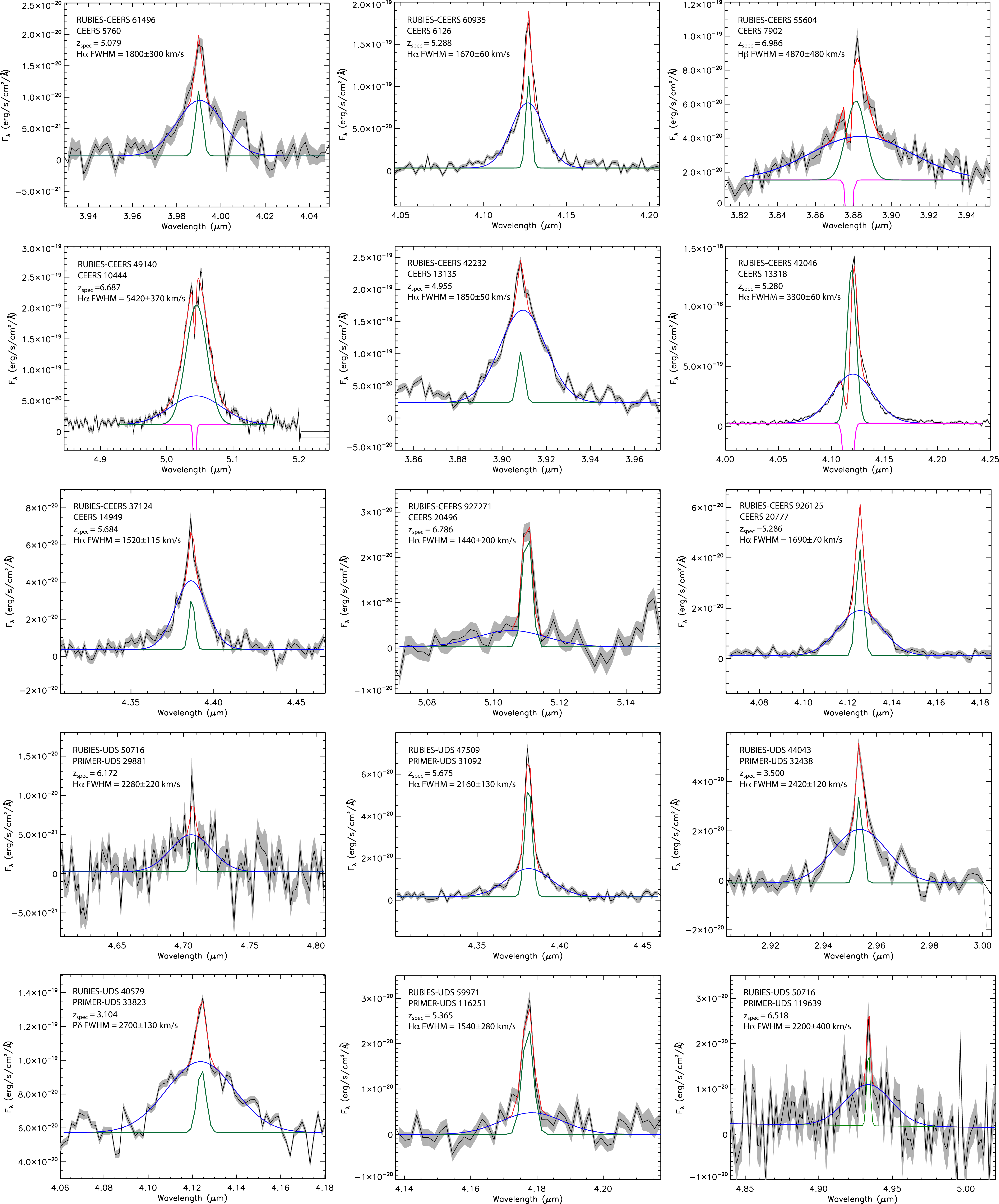}
\caption{NIRSpec spectra with uncertainties (grey shaded region) taken in the G395M grating of 15 LRDs in the EGS and UDS fields that feature broad line emission detected at a ${\rm SNR}>3$. Green lines show the best-fit Gaussian for the narrow emission line component, blue lines show the best-fit broad component, and red lines show the best overall (narrow plus broad) fit the emission line.  The \Hb~emission line is shown for source CEERS 7902 and the \Pd~line for PRIMER-UDS 33823. The \Ha~emission line is shown for all other sources. The FWHM of the broad component (corrected for instrument broadening) is shown in the upper left of each panel.}
\label{fig:BLfits}
\end{figure*}

For PRIMER-COS 3866, it may simply be the fact that it is the brightest LRD in our sample; it is nearly three magnitudes brighter in F444W than the next brightest LRD.  
Using the optical bolometric correction of \citet{Duras20} and assuming that the rest-frame 4400\AA~continuum emission from the LRDs is dominated by AGN emission, PRIMER-COS 3866 has the highest inferred bolometric luminosity of any LRD in our sample.

However,  JADES 21925 is of comparable brightness to several other LRDs in the GOODS-South field that are undetected.  Our X-ray stacking analysis places a $2\sigma$ upper limit on the average flux of these sources as $2.3\times10^{-17}$ erg s$^{-1}$ cm$^{-1}$ in the 2-10 keV band.  
Comparing this flux with what we would expect given their bolometric luminosities and the X-ray bolometric correction from \citet{Duras20}, we find that the X-ray emission of the undetected LRDs is on average 15 times weaker than expected. Such a deficit would require Compton thick levels of obscuration (i.e., $N_{H} > 10^{24} $ cm$^{-2}$).
 For comparison, the predicted X-ray luminosity of JADES 21925 based on its obscuration-corrected, 4400\AA~continuum luminosity is in excellent ($<0.17$ dex) agreement with the X-ray luminosity that we measure for the source.  If most of the LRDs are indeed obscured at the Compton-thick level, then JADES 21925, which has a best-fit column density of $\log\,(N_{\rm H}/{\rm \,cm}^{-2})=22.72^{+0.13}_{-16}$, may be X-ray detected because it is among the least obscured LRDs in our sample.

Alternatively, it has been proposed that the LRD population may be intrinsically X-ray weak due to super-Eddington accretion\citep{Pacucci_Narayan_2024, Volonteri24, Lambrides24, Inayoshi_2024}, which would help explain the high optical-to-X-ray flux ratios observed in LRDs without the need for heavy obscuration.  Interestingly, \cite{Pacucci_Narayan_2024} argued that Chandra's ability to detect this population of faint AGN is strongly dependent on the viewing angle relative to the collimated jets of the central SMBH.  For a jet with random spatial orientation, the probability of observing it within a narrow cone of $10^\circ$ inclination from the pole is $\sim 1.5\%$, based on geometric considerations.  This aligns with the observed detection rate of LRDs in the present sample, suggesting that Chandra may predominantly identify those LRDs aligned nearly along our line of sight to the jet, in a blazar-like configuration.

\subsection{Broad-Line Detections} \label{sec:BLAGN} 

A high fraction of LRDs have been shown to exhibit broad line emission \citep[e.g.,][]{Matthee_2023, Greene_2023}, confirming the presence of an AGN in these sources.  Among our sample, 11 sources were previously reported to have broad \Ha~emission.  These are CEERS 82815 from \citet[][their ID 746]{Kocevski23b}, CEERS 69459 from \citet[][their ID 672]{Harikane23}, JADES 12068 from \citet[][their ID 10013704]{Maiolino23}, UNCOVER 3989, 4535, 9358, 9497, 9904, 25119, and 29255 from \citet[][L23 IDs 30782, 28343, 8798, 8296, 6430, 20080, and 13556]{Greene_2023}, and PRIMER-UDS 33823 from \citet[][their ID RUBIES-BLAGN-1]{Wang24}. These sources are highlighted as cyan diamonds in Figures \ref{fig:slopes} and \ref{fig:mag-size}.  

\begin{figure*}[h!t!]
\centering
\includegraphics[width=\linewidth]{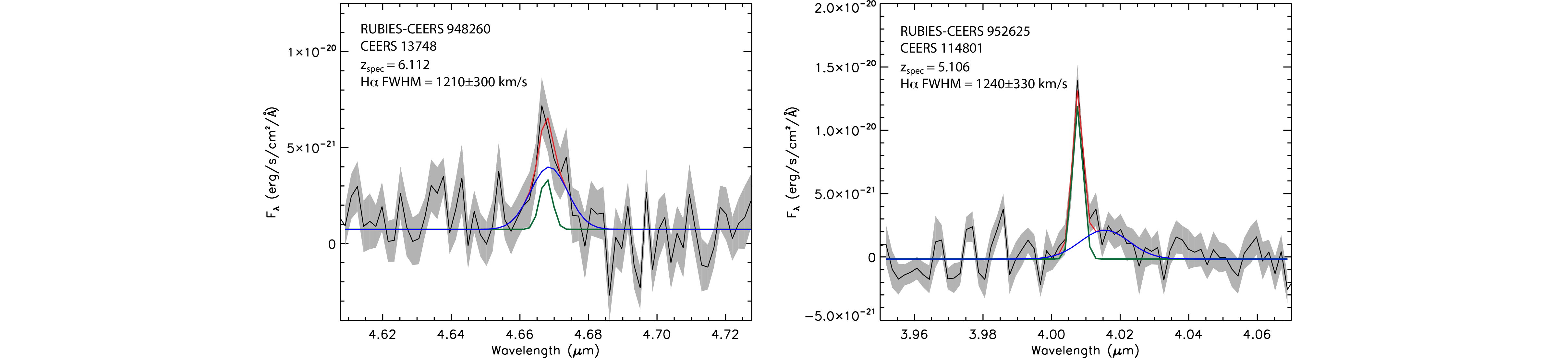}
\caption{NIRSpec spectra with uncertainties (grey shaded region) taken in the G395M grating of 2 LRDs that feature broad line emission detected at a ${\rm SNR}<3$. Model lines are the same as described in Figure \ref{fig:BLfits}.}
\label{fig:BLfits2}
\end{figure*}

\begin{table*}
\renewcommand\thetable{5} 
\caption{Properties of LRDs in our sample with new broad line detections}\vspace{0mm}
\begin{center}
\vspace{-0.25in}
\scriptsize
\begin{tabular}{ccccccccccc}
\hline
\hline
ID & RUBIES ID & RA & Dec & $z_{\rm spec}$ & Line & FWHM$_{\rm broad}$ & $\log\,M_{\rm BH}/{\rm \msun}$ & $\log\,M_{\rm BH}/{\rm \msun}$ & $A_{\rm V}$  & $\log\,M_{\rm *}/{\rm \msun}$\\
  & & (J2000) & (J2000) & & & (km s$^{-1}$) & ($A_{\rm V}=0$) & (Dust Corr.) & & \\
\hline 
    CEERS 5760 & RUBIES-CEERS 61496 & 214.972445 & 52.962196 & 5.079  & \Ha  & $1800\pm300$  & $6.72\pm0.17$ & $7.04\pm0.17$ & 1.8  & $8.53\pm0.33$ \\
    CEERS 6126 & RUBIES-CEERS 60935 & 214.923377 & 52.925588 & 5.288  & \Ha  & $1670\pm60 $  & $7.18\pm0.03$ & $7.63\pm0.03$ & 2.5  & $8.80\pm0.26$ \\
    CEERS 7902 & RUBIES-CEERS 55604 & 214.983037 & 52.956006 & 6.986  & \Hb  & $4860\pm480$  & $8.36\pm0.10$ & $9.31\pm0.10$ & 3.4 & $10.10\pm0.22$ \\
   CEERS 10444 & RUBIES-CEERS 49140 & 214.892248 & 52.877406 & 6.687  & \Ha  & $5420\pm370$  & $8.57\pm0.07$ & $9.11\pm0.07$ & 3.0 & $9.47\pm0.38$ \\
   CEERS 13135 & RUBIES-CEERS 42232 & 214.886801 & 52.855376 & 4.955  & \Ha  & $1850\pm50 $  & $7.39\pm0.03$ & $7.83\pm0.03$ & 2.4 & $8.88 \pm0.32$ \\
   CEERS 13318 & RUBIES-CEERS 42046 & 214.795367 & 52.788848 & 5.280  & \Ha  & $3300\pm60 $  & $8.34\pm0.03$ & $8.88\pm0.03$ & 3.0  & $9.73\pm0.19$ \\
   CEERS 14949 & RUBIES-CEERS 37124 & 214.990983 & 52.916523 & 5.684  & \Ha  & $1520\pm115$  & $6.95\pm0.08$ & $7.46\pm0.08$ & 2.8  & $8.97\pm0.30$ \\
   CEERS 20496 & RUBIES-CEERS 927271 & 215.078257 & 52.948504 & 6.786  & \Ha  & $1410\pm200 $  & $6.45\pm0.18$ & $6.97\pm0.18$ & 2.9 & $8.85\pm0.37$ \\
   CEERS 20777 & RUBIES-CEERS 926125 & 215.137064 & 52.988557 & 5.286  & \Ha  & $1690\pm70 $  & $6.84\pm0.04$ & $7.32\pm0.04$ & 2.7  & $8.80\pm0.25$ \\
    PRIMER-UDS 29881 & RUBIES-UDS 50716 & 34.313132 & -5.226765 & 6.170  & \Ha         & $2280\pm220$ &  $6.98\pm0.11$ & $7.52\pm0.11$ & 3.0  & $8.35\pm0.26$ \\
    PRIMER-UDS 31092 & RUBIES-UDS 47509 & 34.264581 & -5.232544 & 5.675 & \Ha         & $2160\pm130$ &  $7.10\pm0.06$ & $7.45\pm0.06$ & 1.9 & $9.35\pm0.20$ \\
    PRIMER-UDS 32438 & RUBIES-UDS 44043 & 34.241809 & -5.239401 & 3.500 & \Ha         & $2420\pm120$  & $6.98\pm0.05$ & $7.36\pm0.05$ & 2.1 & $7.97\pm0.34$ \\
    PRIMER-UDS 33823 & RUBIES-UDS 40579 & 34.244190 & -5.245834 & 3.103 & Pa$\delta$  & $2700\pm130$  & $8.16\pm0.04$ & $8.29\pm0.04$ & 4.0 & $8.78\pm0.39$ \\
   PRIMER-UDS 116251 & RUBIES-UDS 59971 & 34.260537 & -5.209120 & 5.365 & \Ha         & $1540\pm280$  & $6.44\pm0.20$ & $6.83\pm0.20$ & 2.2 & $8.31\pm0.27$ \\
   PRIMER-UDS 119639 & RUBIES-UDS 63166 & 34.312143 & -5.202546 & 6.518 & \Ha         & $2200\pm400$  & $7.09\pm0.17$ & $7.82\pm0.17$ & 4.0  & $8.24\pm0.41$ \\
 \hline 
\end{tabular}
\normalsize
\end{center}
\label{tab:MBH_results}
\vspace{-0.15in}
\tablecomments{All line widths are reported after correcting for instrument broadening.  PRIMER-UDS 33823 is the same source as RUBIES-BLAGN-1 presented in \cite{Wang24}.}
\end{table*}

We cross-matched our primary sample with the publicly available NIRSpec data in our target fields and have identified 18 sources in the EGS field and six sources in the UDS field that were recently observed by the RUBIES program (GO-4233;  PI: A. de Graaff).  We find evidence for broad emission lines detected with a ${\rm SNR}>3$ in nine CEERS sources (5760, 6126, 7902, 10444, 13135, 13318, 14949, 20496, 20777) and six PRIMER-UDS sources (31092, 32438, 33823, 116251, and 119639).  We also detect tentative broad emission with ${\rm SNR}<3$ in two additional CEERS sources (13748 and 114801). In addition, one source from the sample of B23 (ID 48777) that was removed from our primary sample for having a UV slope of $\beta_{\rm UV}<-2.8$ was also observed and confirmed to be a brown dwarf. 

The G395M/F290LP spectra of the sources that show a broad emission feature, in regions near the broad component, is shown in Figures \ref{fig:BLfits} and \ref{fig:BLfits2}, along with our best-fit two component (narrow plus broad) emission line model for each line.  For these fits, we employ a Levenberg-Marquardt least-squares method described in \citet{Kocevski23b}, where each line is fit with two Gaussians: one narrow with width $\sigma<350$~km~s$^{-1}$ and one broad with width $\sigma>350$~km~s$^{-1}$.   The line centers, widths, and fluxes are all free parameters.  The broad-line widths we measure are listed in Table \ref{tab:MBH_results} and range from $\sim1400$ to 5400 km s$^{-1}$.  The continuum slopes and magnitude distribution of these sources can be seen in Figures \ref{fig:slopes} and \ref{fig:mag-size}.

Of the remaining seven sources observed, we find the emission lines of four are best-fit using a narrow component only (CEERS 2520, 7872, 9083, and 12833), while three sources (CEERS 18850, 99879, and 111399) show only weak \Ha~or \Hb~emission in their G395M/F290LP spectra and it remains inconclusive whether they require a broad component.  These three sources are the faintest of the 24 observed, all with F444W$>26.5$.  The full G395M/F290LP spectra of the 24 LRDs observed by the RUBIES survey are shown in Appendix A.

To determine how the LRDs with confirmed broad-line detections compare to our full LRD sample, we plot the redshift, $\beta_{\rm opt}$, and $M_{\rm UV}$ distributions of both samples in Figures \ref{fig:mag-slope-blagn} and \ref{fig:redshift-MUV-blagn}.  We find that the broad-line LRDs span the full range of $\beta_{\rm opt}$ slopes found in our parent sample and are not limited to the reddest sources.  In terms of their $M_{\rm UV}$, we find the broad-line LRDs are skewed toward brighter sources, having a median $M_{\rm UV}$ of -18.1 as opposed to -18.5 for the full LRD sample.  The sources which lack broad-line detections are preferentially located at higher redshifts ($z>7$), where \Ha~shifts out of the G395M sensitivity window, and among fainter sources with F444W$>26.5$.  Overall, we find the LRDs with broad-line detections are fairly representative of our full LRD sample.

In summary, 15 LRDs out of a sample of 24 observed show evidence of a broad emission component, or 63\%.  This fraction increases to 71\% (17/24) if we include the two sources with broad lines detected at a lower SNR.  If we only consider sources with F444W$<26.5$, where we could effectively measure line widths, the broad-line detection fraction is 71\% (15/21) excluding the low SNR sources and 81\% (17/21) with them included.

\subsubsection{Blueshifted Absorption Features} \label{sec:BLAGN_abs} 

Interestingly, three sources in our sample show evidence of blue-shifted absorption in their Balmer lines (CEERS 7902, 10444, and 13318).  
Adding an absorption component to our fits, we find that these lines have a FWHM in the range 250-500 km s$^{-1}$ and are blue-shifted by 200-300 km s$^{-1}$ relative to the line center of our best-fit narrow component.  Blue-shifted \HeI~$\lambda 1.083 \mu$m absorption was also noted in PRIMER-UDS 33823 by \citet{Wang24}, meaning that 27\% (4/15) of our sample of broad-line LRDs show such features. \cite{Matthee_2023} previously reported the presence of \Ha~absorption lines in two LRDs with broad-line detections, raising the possibility that such features may be common in the spectra of these faint, red AGN.

While broad absorption line quasars (BAL QSOs) often show absorption in their rest-frame UV lines, only a handful of quasars are known to exhibit Balmer absorption; see \citet{Schulze18} and references therein.  
The creation of these narrow but prominent Balmer absorption lines on top of their broad components requires a high column densities of neutral hydrogen in the $n=2$ state.
The effective pumping of the short-live $n=2$ state can be achieved through two primary mechanisms: (1) Ly$\alpha$ trapping in media with high column densities 
($N_{\rm H}\sim 10^{19}~{\rm cm}^{-2}$; \citealt{Hall07}), or (2) collisional excitation in dense environments (hydrogen volume density exceeding $n_{\rm H}>10^9~{\rm cm}^{-2}$; 
\citealt{Juodzbalis_2024, Inayoshi_Maiolino_2024}).  This suggests the absorber is located close to the central engine and is exposed to the quasar's ionizing continuum.  
Previous studies place the location beyond the broad-line region but within the dust sublimation radius \citep{Hall07, Zhang15, Schulze18}.  Therefore the detection of blueshifted Balmer absorption in these LRDs suggests the presence of high density, low ionization gas that is outflowing from near the central engine.

\begin{figure*}[t!]
\centering
\includegraphics[width=\linewidth]{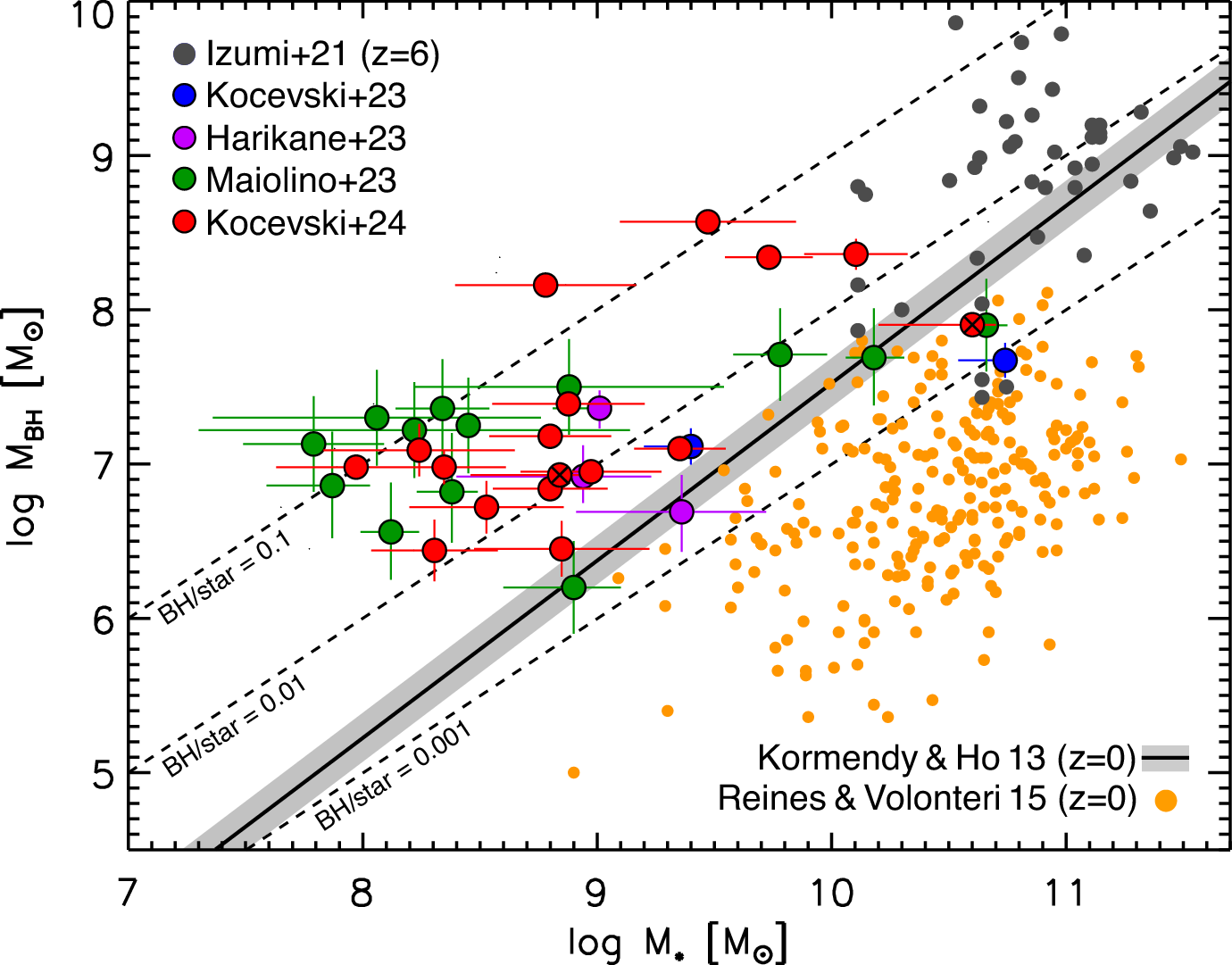}
\caption{The BH mass versus stellar mass relation for our sample of broad-line LRDs (red circles).  The two X-ray detected LRDs in our sample are marked with an X.  Also shown are broad-line AGN recently identified by JWST at $z>4$ drawn from the literature, color coded by the paper in which they were first presented. Dark gray circles show the $z >6$ quasar sample compiled by \cite{Izumi_2021}. Orange circles are the sample of broad-line AGN in the local universe presented in \citet{Reines_Volonteri_2015}, while the gray line is the best-fit local relationship from \citet{Kormendy_Ho_2013}. The diagonal dashed lines represent $M_{\rm BH}/M_\star= 0.1$, $0.01$, and $10^{-3}$.  We find that our broad-line LRDs generally scatter above the local $M_{BH}-M_\star$ relationship. \label{fig:scaling_relationship} }
\end{figure*}

\subsubsection{Black Hole Mass Measurements} \label{sec:BLAGN_MBH} 

Finally, we conclude this section by estimating BH masses for the 15 LRDs with secure broad-line detections.  For this calculation, we follow \citet{Kocevski23b} and make use of the scaling relationships presented in \citet{greene_ho05}.  For sources with \Ha~detections, we use the relationship: 
 \begin{equation} \label{eq: GH05}
 M_{\rm BH} = 2.0 \times 10^6 \left( \frac{L_{\rm H\alpha}}{10^{42}\ {\rm erg\ s^{-1}}}\right)^{0.55}  \left(\frac{{\rm FWHM_{\rm H\alpha}}}{10^3\ {\rm km\ s^{-1}}} \right)^{2.06} M_\odot.
 \end{equation} 
For sources with only \Hb~detections (CEERS 7902), we use the corresponding relationship that employs the \Hb~line width and luminosity:
  \begin{equation} \label{eq: GH05}
   M_{\rm BH} = 3.6 \times 10^6 \left( \frac{L_{\rm H\beta}}{10^{42}\ {\rm erg\ s^{-1}}}\right)^{0.56}  \left(\frac{{\rm FWHM_{\rm H\beta}}}{10^3\ {\rm km\ s^{-1}}} \right)^{2} M_\odot.
 \end{equation} 

In Table \ref{tab:MBH_results}, we list our measured BH masses, both with and without dust corrections to the observed line luminosities.  For this correction we use the $A_{V}$ values derived from our best-fit galaxy plus AGN SED modeling described in \S3.3.  For PRIMER-UDS 33823, we make use of the P$\delta$ line width and luminosity, assuming a P$\delta$-\Ha~line ratio of 46.7, consistent with Case B recombination.  We find that our derived BH mass for this source ($\log M_{\rm BH}/{\rm M}_\odot = 8.26\pm0.04$) is in good agreement with the range of BH masses ($\log M_{\rm BH} / {\rm M}_\odot=7.9-8.6$ ) presented in \cite{Wang24} based on their analysis of multiple broad lines.

In Figure \ref{fig:scaling_relationship} we plot the BH masses of the broad-line LRDs versus our estimate of their host stellar masses, $M_\star$.  Here we estimate $M_\star$ using our SED modeling with {\tt CIGALE v2022.1} , where we assume a hybrid model such that the rest-optical emission of the LRDs originates from reddened AGN emission and the rest-UV light comes from stellar emission from the host galaxy.  The details of this SED modeling are described in \S3.3.  Our derived host masses are listed in Table \ref{tab:MBH_results}.
We find that our LRDs scatter above the local $M_{BH}-M_\star$ relationship seen in massive galaxies at z = 0 \citep[gray line;][]{Kormendy_Ho_2013} and appear overmassive relative to the BH-to-galaxy mass ratio measured for nearby broad-line AGN \citep[orange points;][]{Reines_Volonteri_2015}.  This offset agrees with several previous studies who report similarly overmassive BHs in faint, broad-line AGN identified by JWST at $z > 4$ \citep[e.g.,][]{Kocevski23b, Harikane23,Maiolino23,Pacucci_2023_overmassive, Durodola_2024, Taylor24}.


Although the two X-ray detected LRDs reported in this paper do not have dynamical BH mass measurements, we can estimate the BH-to-galaxy mass ratio for these sources given their X-ray luminosities and assuming they are accreting at the Eddington limit.  The BH and galaxy masses of these sources are shown in Figure \ref{fig:scaling_relationship}.  We find that the BH powering JADES 21925 appears similarly overmassive as our broad-line LRDs, with a BH-to-galaxy mass ratio of $M_{BH}/M_\star = 0.012^{+0.017}_{-0.003}$.  However, PRIMER-COS 3866 is more consistent with the local relationship, having a $M_{BH}/M_\star = 0.002^{+0.003}_{-0.001}$.  We will examine the BH-to-galaxy mass ratio of LRDs and its implications for the evolution of the $M_{BH}-M_\star$ relationship in greater detail in a forthcoming paper.


\begin{figure*}[t!]
\centering
\includegraphics[width=\linewidth]{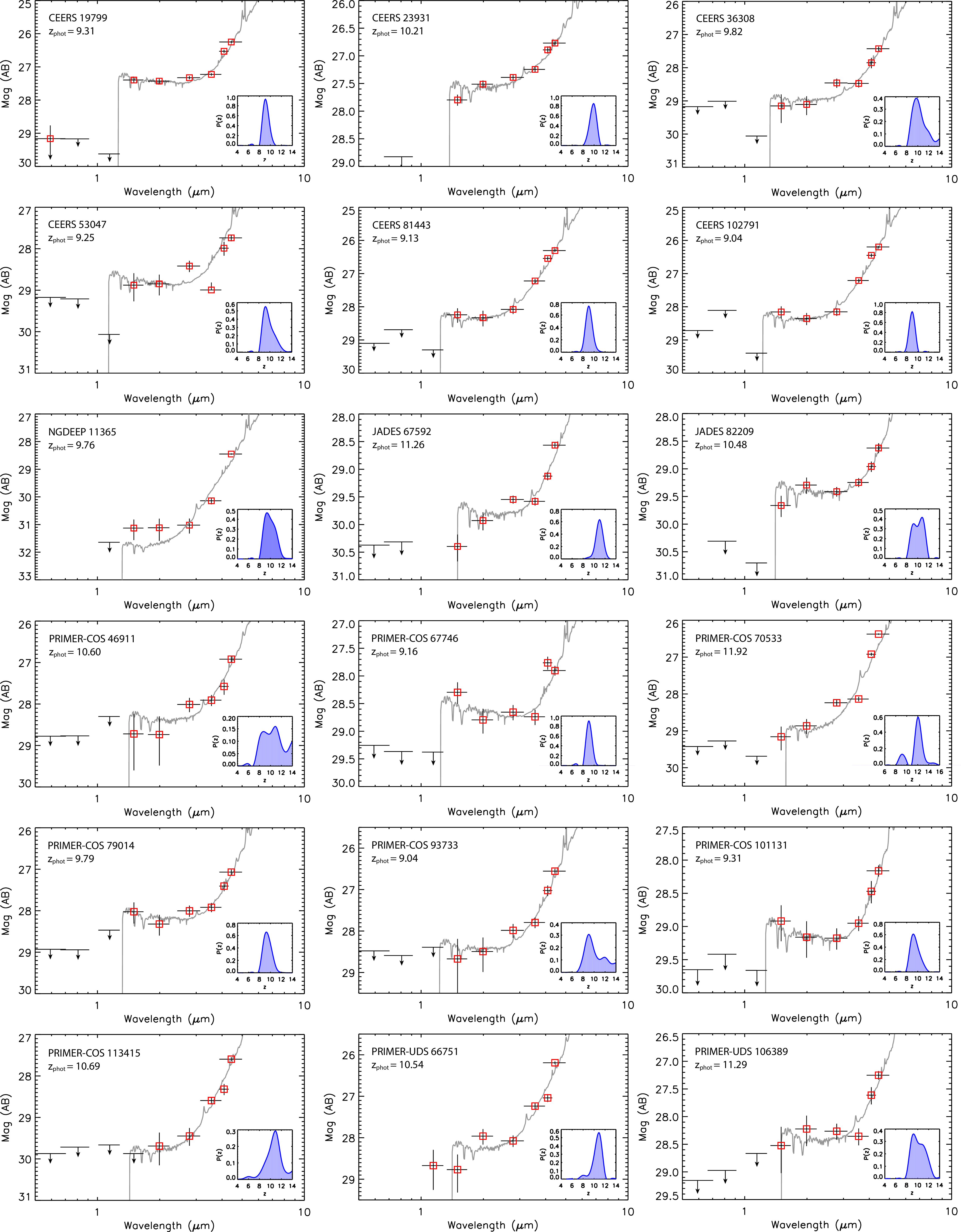}
\caption{Spectral energy distributions for LRDs in our sample that have photometric redshifts of $z > 9$.  The probability density function for the measured photometric redshift of each source is shown in the inset.  $2\sigma$ upper limits are shown for bands with non-detections.}
\label{fig:Highz_seds}
\end{figure*}

\subsection{The $z>9$ LRD Sample} \label{sec:highz} 

In this section, we highlight the highest redshift ($z>9$) LRDs in our sample.  Despite the rapid decline in the redshift distribution above $z\sim8$, we detect 22 LRD at $z>9$.  The SEDs of these sources and their photometric redshift $P(z)$ distributions are shown in Figures \ref{fig:Highz_seds}. 

Four of our LRDs at $z>9$ (CEERS 23931, 36308, 53047, and 80072) were previously identified as high-redshift sources in \citet[their IDs 98518, 92463, 88331, and 1398]{Finkelstein23}.  Four sources at $z>8.5$ (CEERS 2520, 7872, 80438, and 81443) were previously identified as potential high-redshift, massive galaxies in \citet[their IDs 16624, 14924, 21834, and 35300]{Labbe23a}.  

 The highest redshift source in our sample is PRIMER-COS 70533, which has a photometric redshift of 11.92.  However, this source also has a secondary photometric redshift solution at $z=9.2$.  The other three sources with $z>11$ are JADES 67592 and PRIMER-UDS 106389 and 151408, the latter two of which have relatively broad $P(z)$ distributions.  If the redshifts and AGN nature of these sources is confirmed, they would be among the most distant AGN ever discovered, with redshift much higher than the highest-redshift AGN identified in the pre-JWST, which was found at $z=7.6$ \citep{Wang21}.   However, we note that because of the reduced number of filters available for our continuum fitting in the rest-optical at these extreme redshifts, these sources may suffer a greater likelihood of contamination from strong emission lines, which further motivates the need for future deep spectroscopic follow-up of these targets.

\renewcommand{\thefigure}{\arabic{figure} (Cont.)}
\addtocounter{figure}{-1}

\begin{figure*}
\centering
\includegraphics[width=\linewidth]{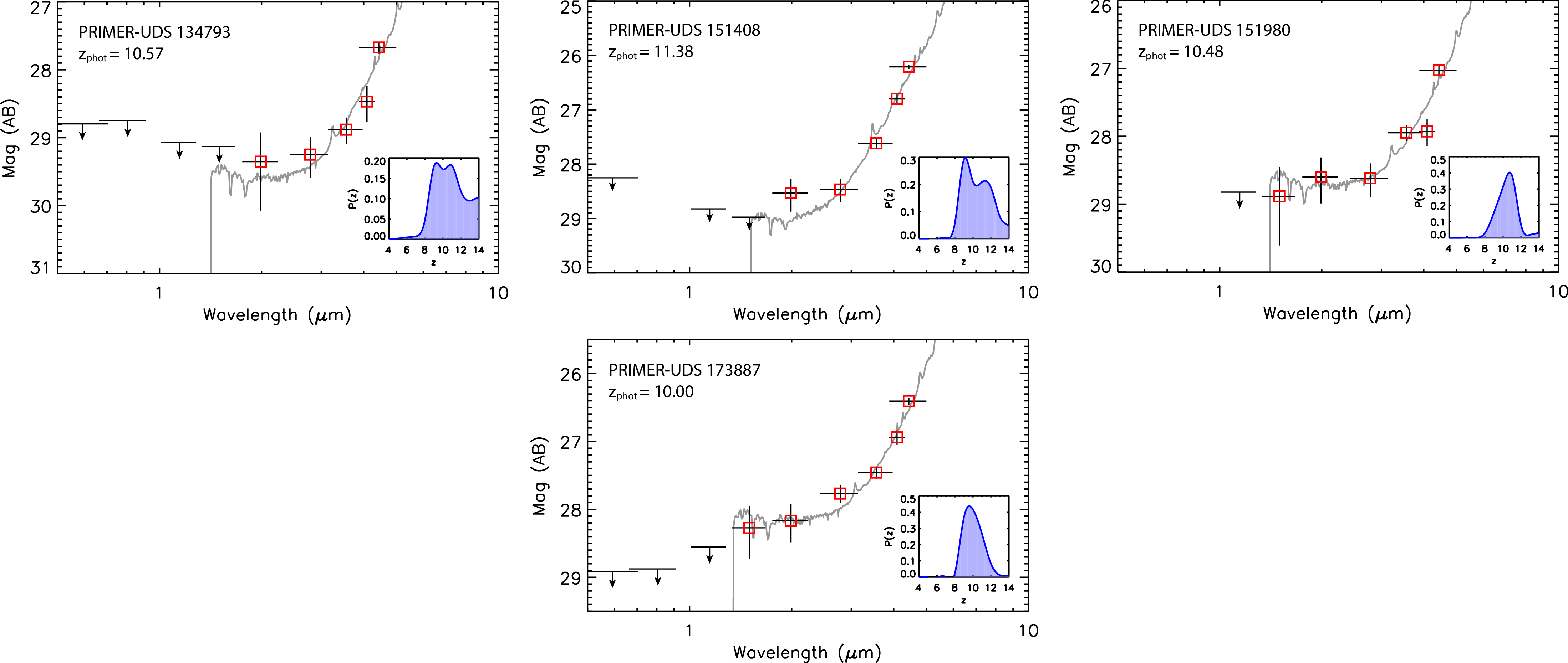}
\caption{Spectral energy distributions for LRDs in our sample that have photometric redshifts of $z > 9$.  The probability density function for the measured photometric redshift of each source is shown in the inset. $2\sigma$ upper limits are shown for bands with non-detections.}
\label{fig:Highz_seds2}
\end{figure*}

\renewcommand{\thefigure}{\arabic{figure}}

\section{Discussion}\label{sec:discussion}

One of our primary findings is that the red compact objects that have come to be known as little red dots appear in large numbers at $z>4$.  Our selection method is designed to pick up lower redshift LRDs that might be missed using a fixed color selection; however, we find few analogs of these sources at $z<4$.  This likely explains why this population was not previously identified using imaging from HST.  

The redshift distribution that we observe for our sample of LRDs may provide insight into the nature of their obscuration and the mechanisms fueling their nuclear activity. The fact that they are largely limited to $z>4$ could be a natural consequence of inside-out growth \citep[e.g.,][]{Carrasco10, vanDokkum14}, where early galaxies experience more compact, centrally concentrated star formation.  The dissipative gas collapse that fuels such activity \citep[e.g.,][]{Dekel14, Tacchella18} may also be responsible for triggering concurrent growth of the central SMBH \citep{Kocevski17, Aird22}. The rapid accumulation of metals in the proto-bulge then provides the reddening we observe.  At later epochs, infalling gas will have higher angular momentum, which results in star formation moving to larger scales.  As less dust is deposited near the AGN, we observe fewer heavily-reddened LRDs at lower redshifts.

In this scenario, the nuclear emission from the central AGN would be reddened by host obscuration as opposed to nuclear obscuration. 
Previous work on red quasars at lower redshifts ($z<0.7$) have found that they are indeed predominantly obscured by dust in their host galaxies, as opposed to, for example, a moderate viewing angle that passes through a dust torus \citep{Kim18_RQSO}.  In addition, \cite{Gilli22} have used deep ALMA observations to show that the ISM column density toward the nucleus of galaxies at $z>6$ can reach Compton-thick levels of obscuration \citep{Gilli22}.  They conclude that 80-90\% of SMBH growth at these early epochs is likely hidden from view due to the ISM of their host galaxies. 

Obscuration from a compact host galaxy would be consistent with the reported high fraction of LRDs that are not detected using MIRI imaging out to $21\mu m$ \citep{Williams23, Pablo24} since galactic-scale dust would produce reprocessed emission that falls beyond the MIRI wavelength coverage.  That said, the mid-infrared emission from the X-ray detected LRDs we have identified at $z\sim3-4$ is well explained using a clumpy torus model, which suggests a fraction of these sources may indeed be obscured by a dusty torus.  It is likely that different scales of obscuration exist in these systems, as has been previously inferred from samples of infrared-selected red quasars \citep[e.g.,][]{Yan19}.  Better constraints on the fraction of LRDs that exhibit hot dust emission may help elucidate the scale of the obscuration in these systems and the prevalence of nuclear-scale dust.  We plan to examine the mid-infrared properties of our sample in a forthcoming paper (Leung et al., in prep).

Another source of obscuration in these systems can be inferred from the presence of Balmer absorption lines in the spectra of several LRDs.  These absorption features may help explain one of the more perplexing properties of LRDs: their lack of X-ray emission.
While we have presented two X-ray bright LRDs, the vast majority of them (including all with broad-line detections) remain undetected even in the deepest \emph{Chandra} observations.  The fraction of LRDs that show Balmer absorption features provides information on the fraction of sight-lines to the central engine that are obscured by high-density, neutral hydrogen clouds, i.e. the covering factor.  The higher fraction that we find relative to that observed in BAL QSOs, for example, suggests an increase in the covering factor among these fainter AGN.  At the gas densities required to produce these Balmer absorption lines, soft x-ray photons will ionize neutral hydrogen and create \Lya~photons via recombination, which subsequently get trapped.  Since most BAL QSOs are X-ray weak \citep{Brandt00, Gallagher01}, we may surmise that the presence of this high-density, absorbing material may also help attenuate the X-ray emission from LRDs.  Coupled with potentially high column densities in the ISM of their host galaxies \citep[e.g.,][]{Gilli22}, this may help explain the low fraction of LRDs that are X-ray detected.

While it remains to be determined what fraction of our full sample are reddened AGN, we find that 17 of 21 newly-observed LRDs where we could effectively measure emission line widths in the G395M/F290LP spectra show evidence for broad-line emission, which is consistent with the high fraction (80\%) reported in \cite{Greene_2023} when brown dwarf contaminants are excluded.  Such a high AGN fraction would have important ramifications for the bright galaxy samples being reported at high redshift.  For example, 11/13 of the massive galaxy candidates originally presented in \citet{Labbe23a} are identified as LRDs using our selection.  If the rest-optical emission of these sources is dominated by AGN emission, as seen in other LRDs, then this may help explain the anomalously high stellar masses reported for these sources. 

The discovery of a new population of AGNs at high redshift naturally begs the question of whether this population could significantly contribute to the reionization photon budget.  While the declining abundance of UV-bright quasars implies that such objects are not contributors at z > 6 \citep[e.g.,][]{Hopkins07}, the uncertain abundance of UV-faint AGN leaves open the door for a modest contribution by such sources \citep[e.g.][]{Giallongo15, Finkelstein19}.  While the high bolometric luminosity inferred for LRDs 
\citep[$L_{\rm bol}\sim10^{43}-10^{46}$ erg/s; e.g.,][]{Labbe23, Kokorev24} implies the AGNs may be producing large quantities of ionizing photons, the high dust reddening measured in these sources \citep[$A_{\rm V} \geq 3$;][]{Barro23, Labbe23} strongly limits the escape of ionizing photons.  This is evidenced by the weak UV emission from these galaxies.  While it remains unclear whether the UV emission is dominated by unobscured stellar light or scatted AGN light, the abundance of LRDs is $\sim100\times$ lower than continuum-selected star-forming galaxies (see Figure 8).  Thus even if the escaping UV light is AGN-dominated, it is insignificant compared to the known star-forming galaxy population at these redshifts.

\section{Conclusions}

We present the largest sample of little red dots compiled to date using JWST data from the CEERS, PRIMER, JADES, UNCOVER, and NGDEEP surveys.  Our sample contains 341 sources spanning the redshift range $2<z<11$ in a total area covering 587.8 arcmin$^{2}$.  These sources are selected to have red colors in the rest-frame optical and blue colors in the rest-frame UV, coupled with a compact morphology.   While previous studies used color indices measured in a single combination of bands (i.e., F277-F444W) to identify LRDs, we perform our search by fitting the UV and optical continuum slope of sources using multiple bands blueward and redward of the Balmer break.  The bands used for these fits shift as a function of redshift to ensure the same rest-frame emission is sampled for each source.  This allows us to identify LRDs over a wider redshift range than a fixed color selection and is less susceptible to contamination from galaxies with strong breaks that otherwise lack a rising red continuum.

Using our sample, we find:
\begin{itemize}
\item LRDs emerge in large numbers at $z\sim8$ and then experience a rapid decline in their number density at $z\sim4.5$.  This redshift distribution may reflect that the fueling and obscuration of these sources is tied to the dissipative gas collapse that drives the inside-out growth of galaxies during this epoch. In this scenario, the decline in the number of LRDs at later times would be due to star formation moving to larger scales, resulting in less dust being deposited near the AGN and a possible drop in their duty cycle.

\item LRDs are 2-3 dex more numerous at $z\sim5-7$ than would be expected based on extrapolations of the quasar luminosity function measured from the ground.  However, the number density of LRDs at $z\sim5$ is only 0.6 dex higher than that of lower-luminosity X-ray AGN identified at the same redshift.  This offset increases to a full dex at $z\sim7$ relative to color-selected AGN.  At this redshift, LRDs constitute $\sim3\%$ of the overall galaxy population at $M_{\rm UV}=-20$.

\item We identify the first LRDs detected at X-ray wavelengths.  An X-ray spectral analysis of these two sources confirms that they are moderately obscured, with equivalent neutral Hydrogen column densities of $\log\,(N_{\rm H}/{\rm cm}^{-2}) =$ $23.3^{+0.4}_{-1.3}$ and $22.72^{+0.13}_{-0.16}$. An analysis of their SEDs and morphologies suggests the rest-frame optical emission from both sources is dominated by light from the reddened AGN, while their rest-frame UV emission originates from their host galaxies. 

\item We present follow-up spectroscopy of 17 LRDs in our sample that show broad emission features with line widths of $\sim1400$ to 5400 km s$^{-1}$, consistent with AGN activity.  We measure their black hole masses to be in the range $M_{\rm BH} = 10^{6-8}$ M$_\odot$, which is 1-2 dex below that of bright quasars at similar redshifts. 

\item The confirmed AGN fraction of our sample is 71\% among sources with F444W$<26.5$, where we could effectively measure emission line widths in the G395M/F290LP spectra.  This fraction increases to 81\% if we include sources whose broad emission lines are detected with a $SNR<3$.

\item A relatively high fraction (24\%) of LRDs in our sample with broad emission lines also show narrow, blue-shifted Balmer or \HeI~absorption features in their spectra, suggesting a prevalence of outflows in these sources.  We propose that a high covering factor of high-density, neutral hydrogen gas, coupled with potentially high column densities in the ISM of their host galaxies, may help explain the relatively weak X-ray emission from LRDs.

\item We find that our LRDs scatter above the local $M_{BH}-M_\star$ relationship seen in nearby massive galaxies and appear overmassive relative to the BH-to-galaxy mass ratio measured for low-redshift, broad-line AGN.

\end{itemize}

While much remains to be determined about the nature of LRDs, the prevalence of broad emission lines in their spectra suggests this population is shedding light on a phase of obscured black hole growth in the early universe that was largely undetected prior to the JWST era.  Forthcoming mid-infrared imaging and spectroscopic follow-up of these sources will soon better constrain the number density of these faint AGN and shed light on the nature of the obscuring medium in these systems.

\appendix
\section{NIRSpec Spectra of Newly Observed LRDs}

In this appendix, we present the full G395M/F290LP spectra of the 24 LRDs from our sample in the EGS and UDS fields that were recently observed by the RUBIES program.  These spectra are shown in Figures 18, 19, 20, 21, 22, and 23.

\begin{figure*}
\centering
\includegraphics[width=\linewidth]{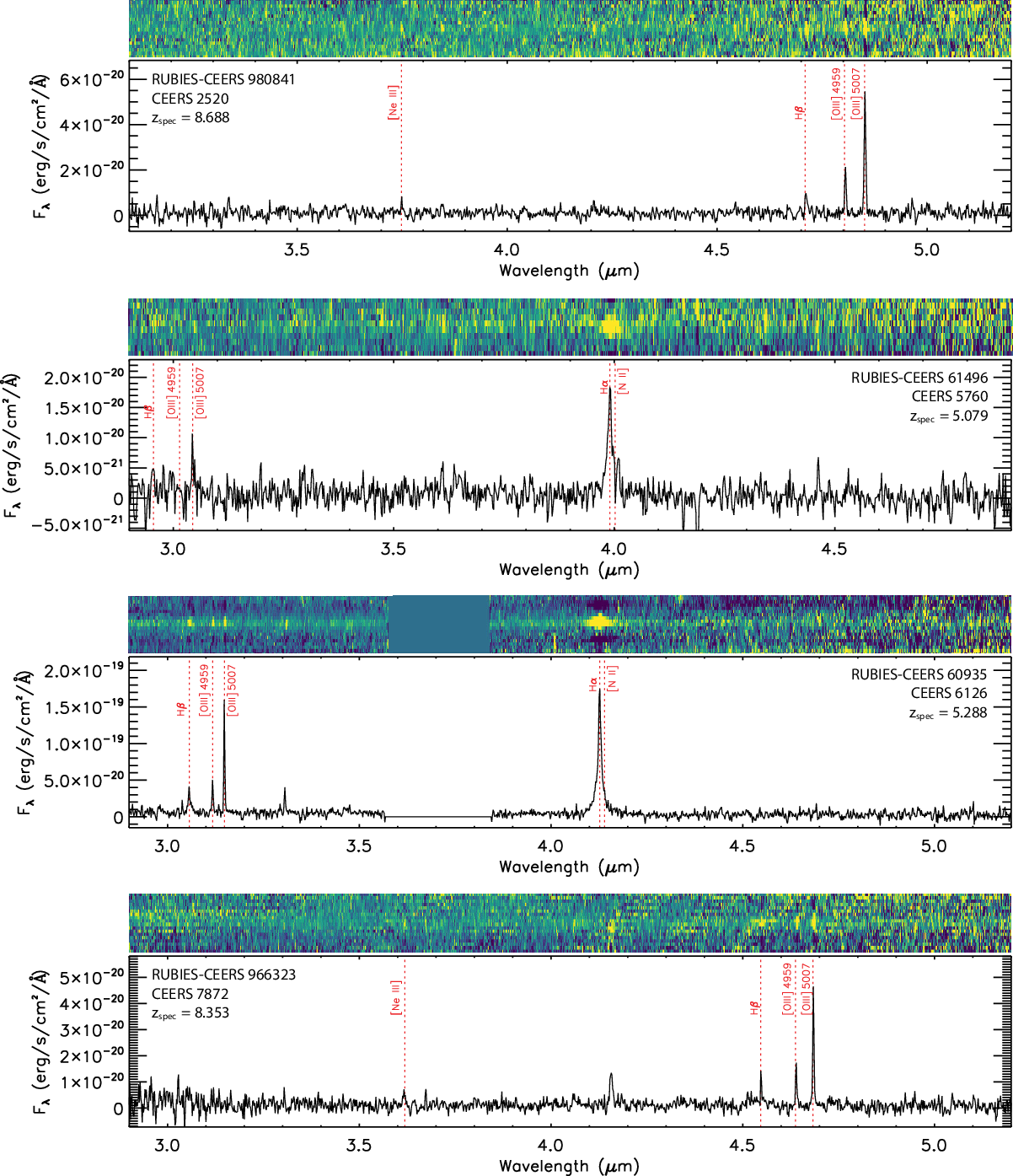}
\caption{NIRSpec spectra from the RUBIES survey of sources CEERS 2520, 5760, 6126, and 7872 taken in the G395M grating. The locations of several prominent emission lines are noted.}
\label{fig:full_spectra1}
\end{figure*}

\begin{figure*}
\centering
\includegraphics[width=\linewidth]{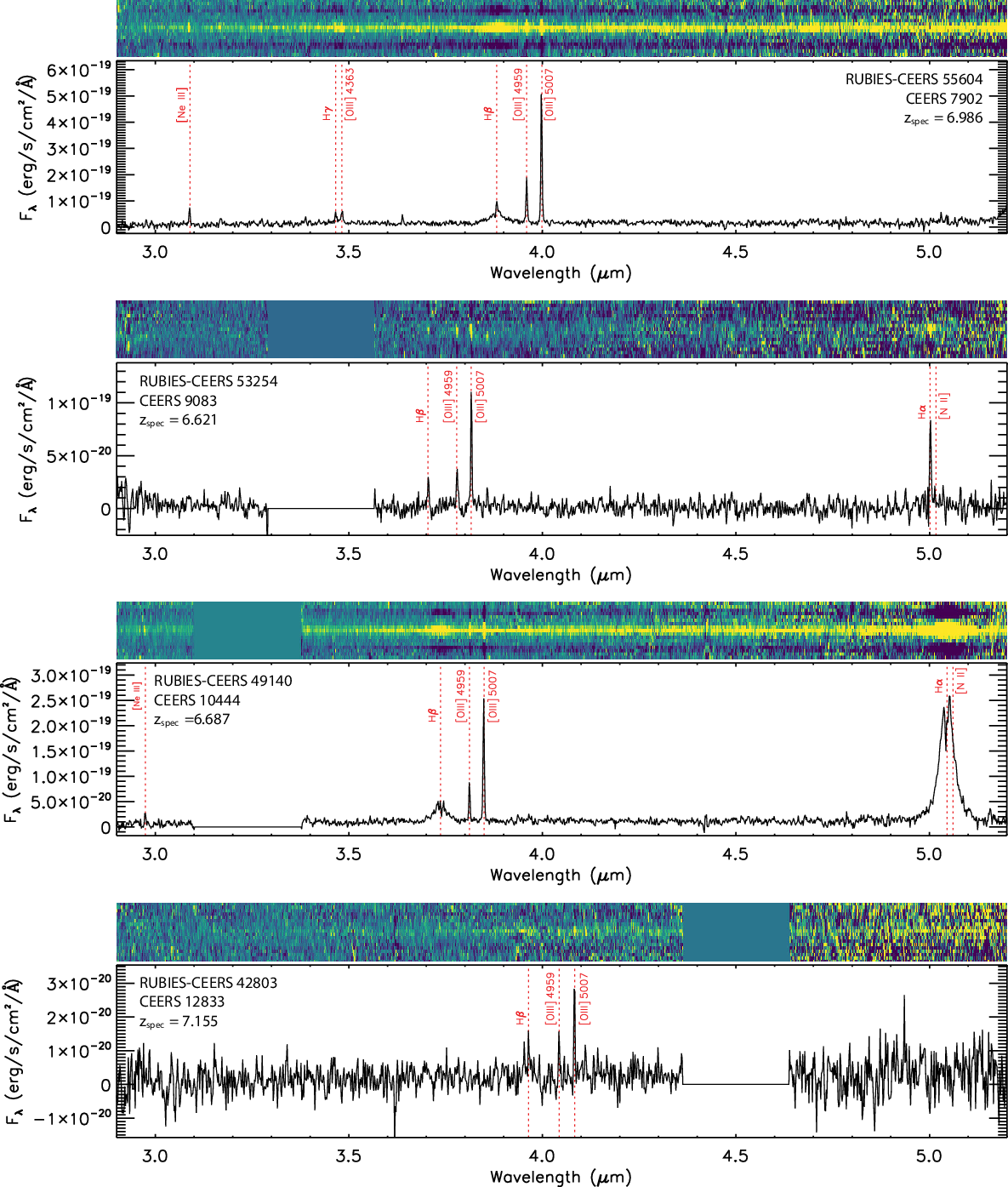}
\caption{NIRSpec spectra from the RUBIES survey of sources CEERS 7902, 9083, 10444, and 12833 taken in the G395M grating. The locations of several prominent emission lines are noted.}
\label{fig:full_spectra1}
\end{figure*}

\begin{figure*}
\centering
\includegraphics[width=\linewidth]{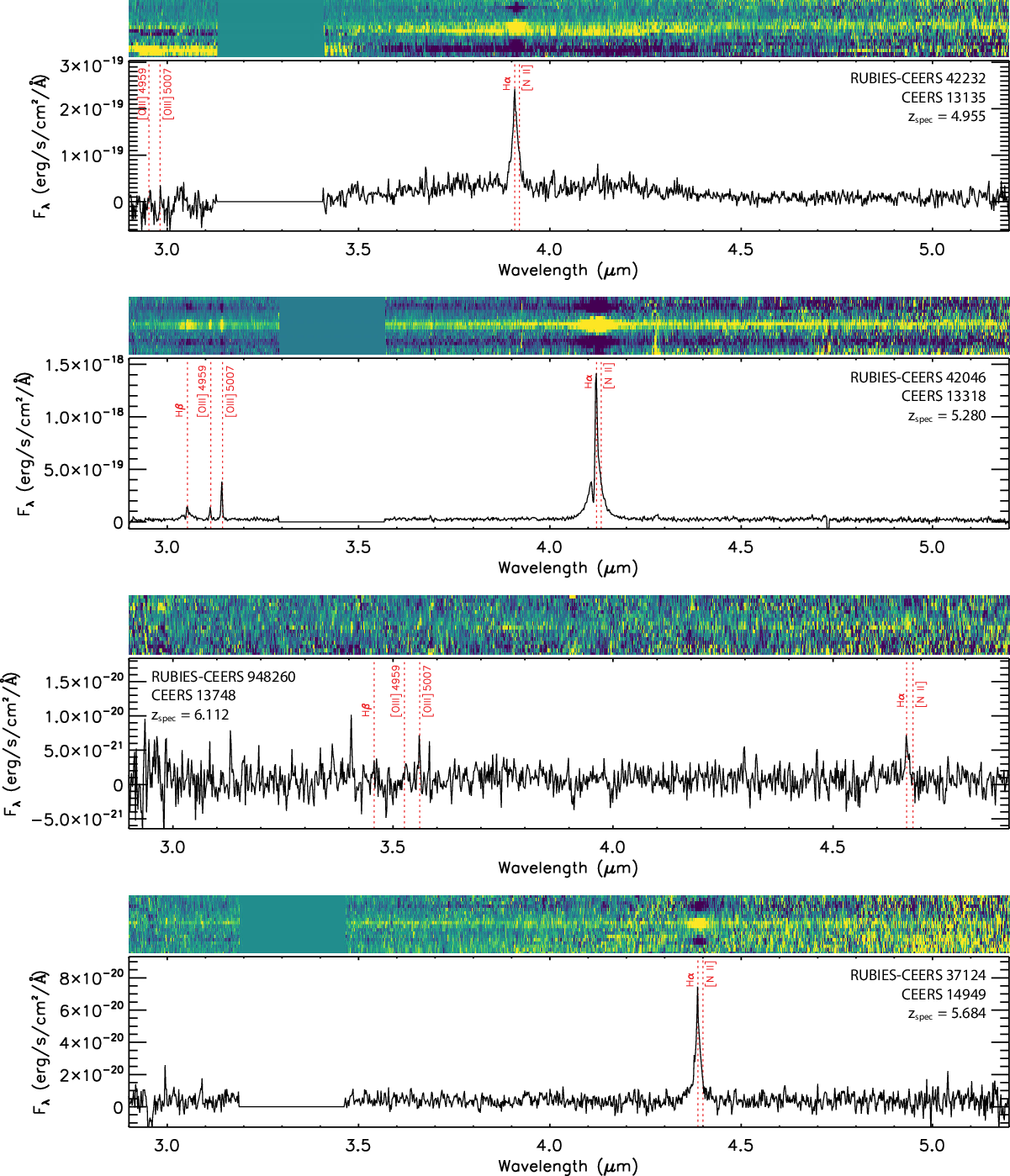}
\caption{NIRSpec spectra from the RUBIES survey of sources CEERS 13135, 13318, 13748, and 14949 taken in the G395M grating. The locations of several prominent emission lines are noted.}
\label{fig:full_spectra1}
\end{figure*}

\begin{figure*}
\centering
\includegraphics[width=\linewidth]{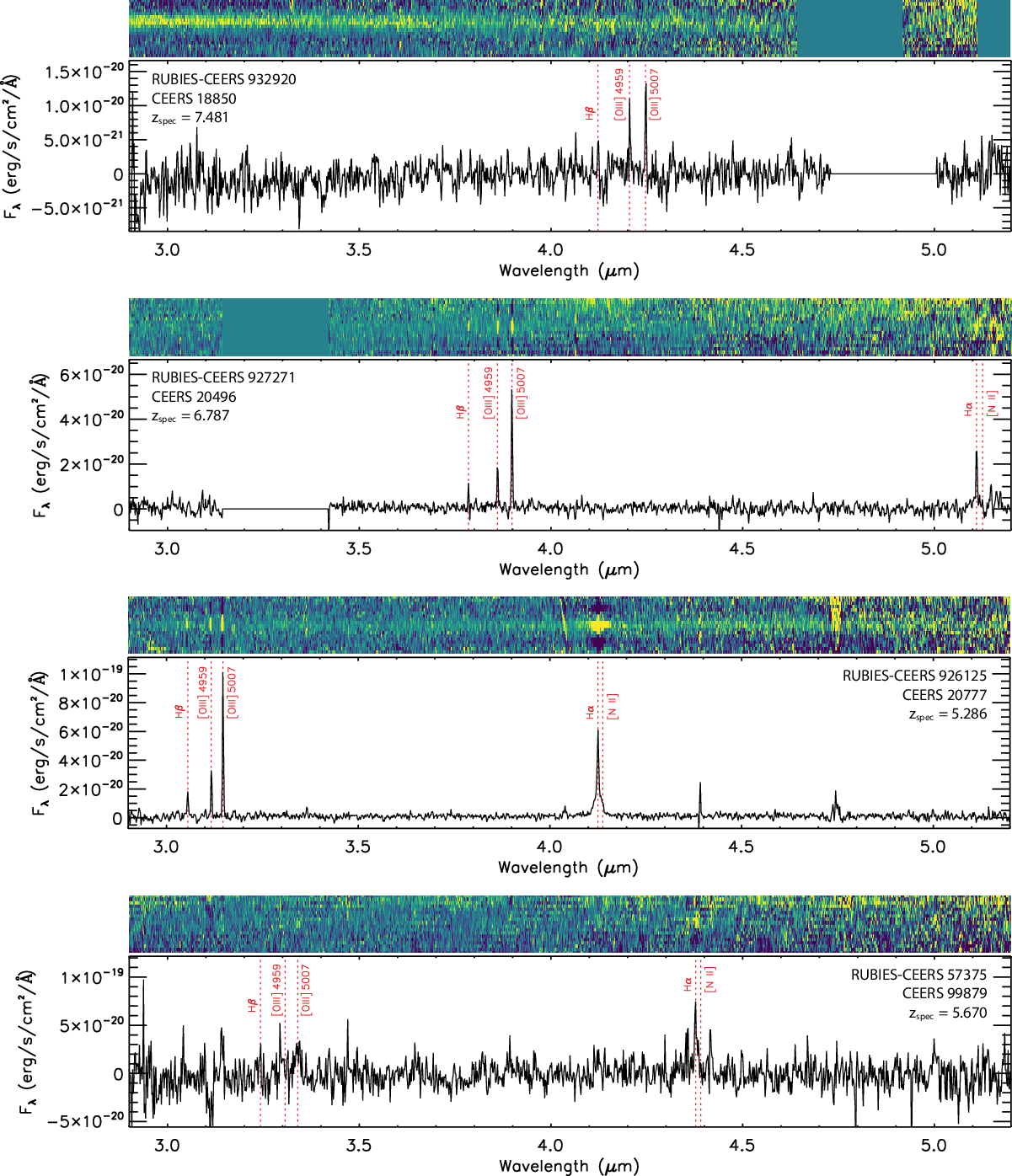}
\caption{NIRSpec spectra from the RUBIES survey of sources CEERS 18850, 20496, 20777, and 99879 taken in the G395M grating. The locations of several prominent emission lines are noted.}
\label{fig:full_spectra1}
\end{figure*}

\begin{figure*}
\centering
\includegraphics[width=\linewidth]{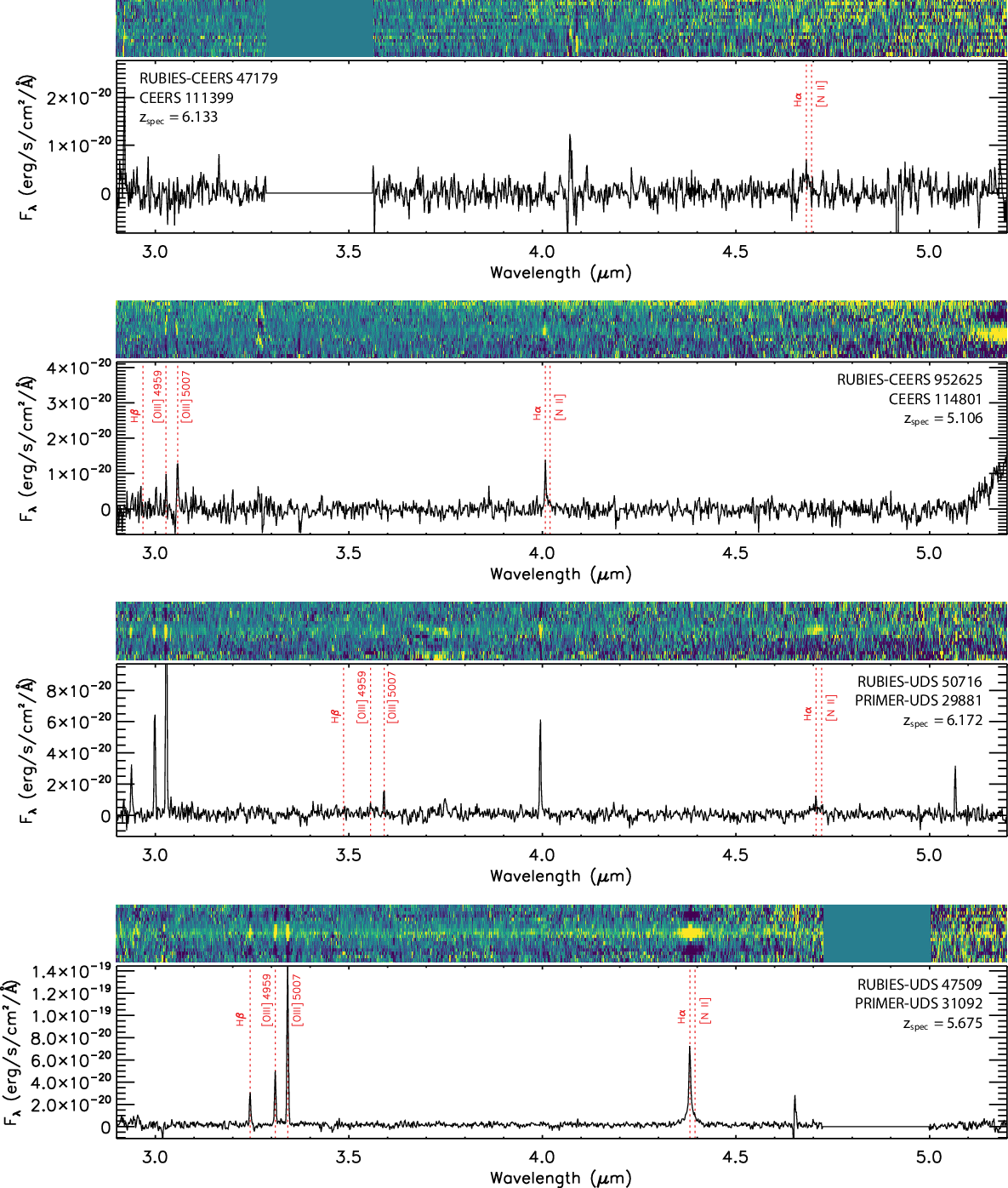}
\caption{NIRSpec spectra from the RUBIES survey of sources CEERS 111399 and 114801, as well as PRIMER-UDS 29881 and 31092 taken in the G395M grating. The locations of several prominent emission lines are noted.}
\label{fig:full_spectra1}
\end{figure*}

\begin{figure*}
\centering
\includegraphics[width=\linewidth]{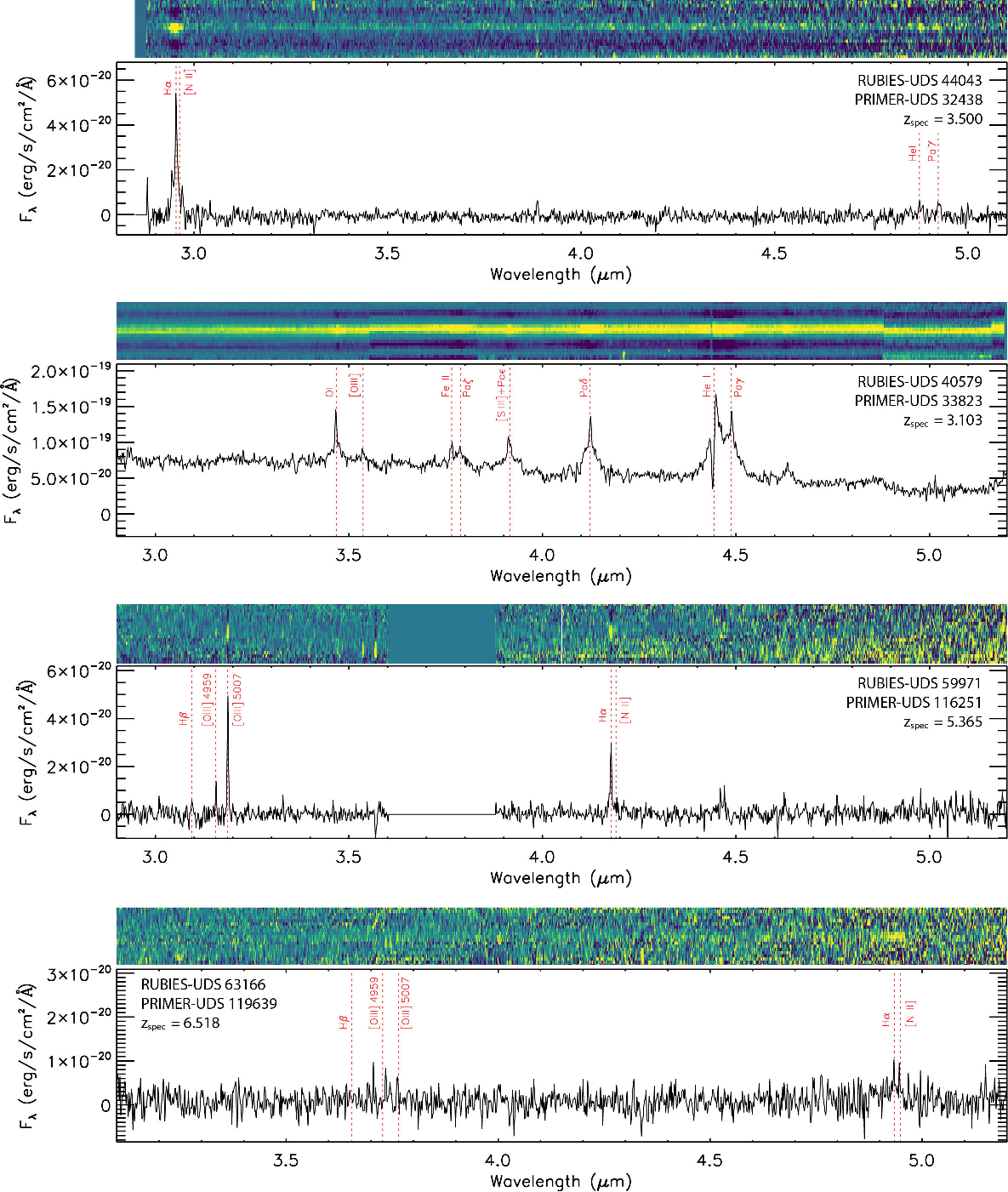}
\caption{NIRSpec spectra from the RUBIES survey of sources PRIMER-UDS 32438, 33823, 116251, and 119639 taken in the G395M grating. The locations of several prominent emission lines are noted.  PRIMER-UDS 33823 is the same source as RUBIES-BLAGN-1 presented in \cite{Wang24}.}
\label{fig:full_spectra1}
\end{figure*}

\section{Line Boosted LRD Candidates}

\begin{figure}
\centering
\includegraphics[width=\linewidth]
{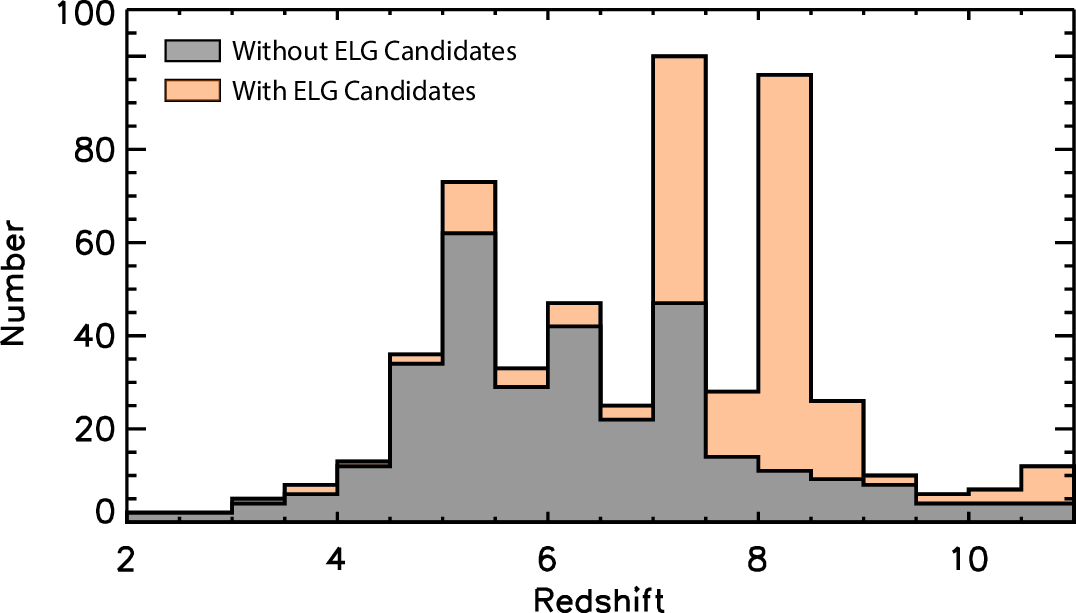}
\caption{The redshift distribution of our LRDs sample when including sources potentially contaminated by emission line boosting (orange).  The redshift distribution of our final sample of LRDs, which excludes such sources is shown in gray.}
\label{fig:redshift_hist_ELGs}
\end{figure}

\begin{figure*}
\centering
\includegraphics[width=\linewidth]{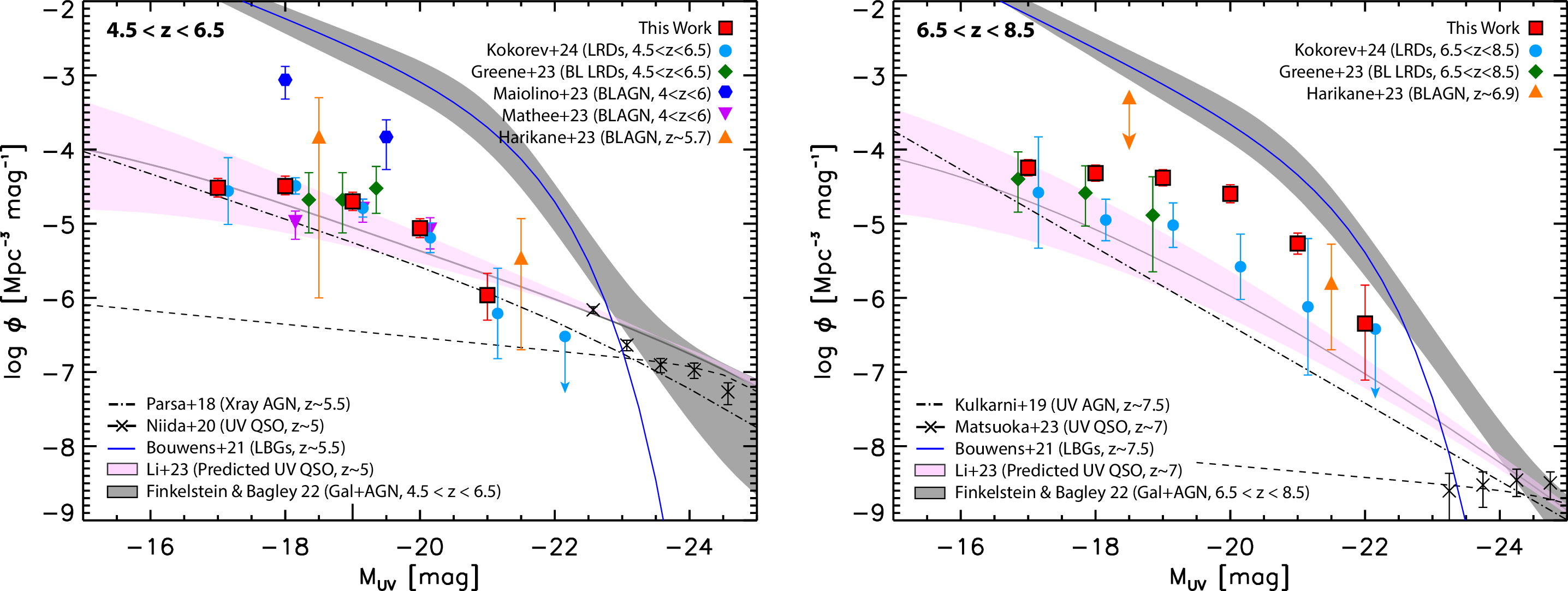}
\caption{The UV luminosity function of our LRD sample when including sources potentially contaminated by emission line boosting, presented in two redshift bins: $4.5<z<6.5$ (\emph{left}) and $6.5<z<8.5$ (\emph{right}). The UV luminosity is measured at rest-frame 1450\AA.  We find good agreement with the luminosity functions inferred from previous photometric and spectroscopic compilations of LRDs at $z\sim5$, however, at $z\sim7$ our measured number density with these sources included is elevated relative to values reported in the literature by \cite{Greene_2023} and \cite{Kokorev24}.}
\label{fig:LF_wELGs}
\end{figure*}

The LRD sample selection outlined in \S3.1 includes cuts meant to remove sources whose $\beta_{\rm opt}$ slope may be boosted due to strong line emission affecting one or more bands.  These cuts eliminate 195 potential LRDs from our parent sample.  In this section, we provide additional details about these potentially line boosted sources
and discuss how our conclusions would change if these sources had not been removed from our LRD sample.  The coordinates, redshifts, and best-fit continuum slopes of these sources are reported in Table \ref{tbl:ELG_sample}.  Figure \ref{fig:redshift_hist_ELGs} shows the redshift distribution of our LRDs when including sources potentially contaminated by line emission.  
Keeping these sources results in a dramatic increase in the number of potential LRDs at $7<z<9$, however the distribution clearly shows signs of strong emission lines entering our selection bands at specific redshifts.  In Figure \ref{fig:LF_wELGs} we show the UV luminosity function we measure with these sources included in our LRD sample.  While we find good agreement with the luminosity functions inferred from previous photometric and spectroscopic compilations of LRDs at $z\sim5$, at $z\sim7$ our measured number density with these sources included is elevated relative to values reported in the literature by \cite{Greene_2023} and \cite{Kokorev24}.  This is likely because the LRDs which were removed from our sample for potential line contamination would also fail the ${\rm F277W-F356W} > 0.6$ color cut imposed by \cite{Kokorev24} due to their relatively flat SEDs in the rest-optical.  In fact, of the 195 potentially line boosted LRDs removed from our sample, only two are included in the sample presented in \cite{Kokorev24}.  Nonetheless, we provide the coordinates of these sources in Table \ref{tbl:ELG_sample} as potential targets for future follow-up observations.

\begin{deluxetable*}{lccccccccccc}[t]
\tablenum{3}
\tablecolumns{7}
\tablecaption{Properties of Little Red Dots cut from our parent sample for potential line boosting \label{tbl:ELG_sample}}
\tablehead{
 \colhead{ID} & \colhead{RA} & \colhead{Dec} & \colhead{$z_{\rm best}$} &  \colhead{$z_{\rm flag}$} & \colhead{$\beta_{\rm UV}$} & \colhead{$\beta_{\rm opt}$} & \colhead{$m_{\rm F444W}$}& \colhead{$M_{\rm UV}$} & \\ 
 \colhead{} & \colhead{(J2000)} & \colhead{(J2000)} & \colhead{} & \colhead{} & \colhead{} & \colhead{} & \colhead{(AB mag)} & \colhead{(AB mag)}  }
\startdata
          CEERS 709 & 214.771840 & 52.850962 & 8.17 & 1 & $-2.14\pm0.19$ & $ 0.14\pm0.36$ & 26.69 & -19.88 \\
          CEERS 7602 & 215.008653 & 52.976511 & 6.01 & 1 & $-2.03\pm0.13$ & $ 0.21\pm0.06$ & 25.09 & -20.10 \\
         CEERS 10262 & 214.938642 & 52.911754 & 8.98 & 1 & $-2.42\pm0.10$ & $ 0.26\pm0.19$ & 26.37 & -20.69 \\
         CEERS 10807 & 214.968700 & 52.929653 & 8.92 & 1 & $-2.48\pm0.15$ & $ 0.21\pm0.31$ & 26.47 & -20.72 \\
         CEERS 14877 & 214.882997 & 52.840416 & 8.05 & 1 & $-2.17\pm0.08$ & $ 1.59\pm0.15$ & 25.61 & -20.75 \\
         CEERS 22451 & 215.003051 & 52.885675 & 7.42 & 1 & $-2.15\pm0.11$ & $-0.02\pm0.08$ & 26.43 & -19.83 \\
         CEERS 23173 & 215.037195 & 52.906711 & 7.39 & 1 & $-1.87\pm0.10$ & $ 0.32\pm0.07$ & 25.58 & -20.06 \\
         CEERS 26539 & 215.047058 & 52.897477 & 8.32 & 1 & $-2.32\pm0.14$ & $ 0.97\pm0.27$ & 26.93 & -19.59 \\
         CEERS 27802 & 215.089926 & 52.922061 & 8.20 & 1 & $-2.18\pm0.20$ & $ 1.52\pm0.29$ & 26.47 & -19.60 \\
         CEERS 27865 & 215.129877 & 52.949951 & 7.39 & 1 & $-1.96\pm0.17$ & $ 0.31\pm0.10$ & 25.75 & -20.03 \\
\enddata
\tablecomments{$z_{\rm flag}$: 1 = phot redshift, 2 = spect redshift.}
\end{deluxetable*}

\section{Acknowledgments}

We thank the RUBIES team for their work in designing and preparing the NIRSpec observations used in this study.  This work is supported by NASA grants JWST-ERS-01345 and JWST-AR-02446 and based on observations made with the NASA/ESA/CSA James Webb Space Telescope. The data were obtained from the Mikulski Archive for Space Telescopes at the Space Telescope Science Institute, which is operated by the Association of Universities for Research in Astronomy, Inc., under NASA contract NAS 5-03127 for JWST.


\end{document}